\documentclass[times, review, 10pt]{elsarticle}

\usepackage{amssymb}
\usepackage{lipsum}
\usepackage{amsthm}
\usepackage{amsmath}

\newtheorem{example}{Example}

\let\oldexample\example
\let\endoldexample\endexample
\renewenvironment{example}[1][]{%
 \oldexample[#1]\normalfont 
}{\endoldexample}

\usepackage{wasysym}
\usepackage{algorithm}
\usepackage[noend]{algpseudocode}
\usepackage{algpseudocode}

\usepackage{tabularray}
\usepackage{makecell}
\usepackage{multirow}
\usepackage{booktabs}
\usepackage{pifont}
\usepackage{utfsym}
\usepackage{ulem}
\usepackage{natbib}
\setcitestyle{comma}
\setcitestyle{numbers,square}

\usepackage{graphicx}
\usepackage[utf8]{inputenc}
\usepackage{tcolorbox}

\usepackage{url}
\renewcommand{\emph}[1]{\textit{#1}}
\usepackage{graphicx} 
\usepackage{tikz} 
\usepackage{simpleicons}
\usepackage{listings}    
\usepackage[table]{xcolor}      
\usepackage{caption}     

\usepackage{tcolorbox}
\tcbuselibrary{listings,skins}
\tcbuselibrary{breakable}    
\tcbset{
  promptbox/.style={
    colback=gray!10,         
    colframe=black!70,       
    fonttitle=\bfseries,
    title=Prompt,            
    sharp corners,
    boxrule=0.6pt,
    arc=2pt,
    left=6pt, right=6pt, top=4pt, bottom=4pt,
    breakable               
  }
}
\newtcolorbox{PromptBox}[1][]{
  promptbox, #1
}

\makeatletter
\def\ps@pprintTitle{%
    \let\@oddhead\@empty
    \let\@evenhead\@empty
    \def\@oddfoot{\reset@font\hfil\thepage\hfil}
    \let\@evenfoot\@oddfoot
}
\makeatother
\journal{Pattern Recognition}

\begin{document}


\begin{frontmatter}



\title{%
  \begin{minipage}[c]{\linewidth} 
    \begin{minipage}[c]{2.5cm}   
      \includegraphics[width=2.2cm, height=2.2cm]{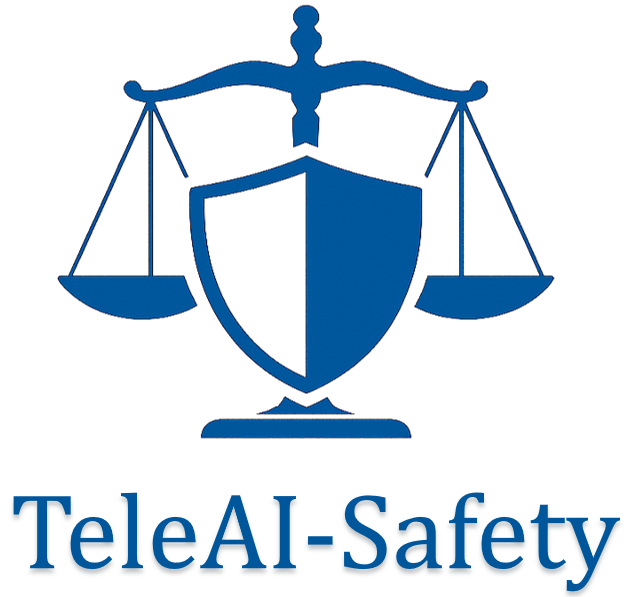}
    \end{minipage}%
    \begin{minipage}[c]{\dimexpr\linewidth-3cm\relax} 
      \textbf{TeleAI-Safety: A comprehensive LLM jailbreaking benchmark towards attacks, defenses, and evaluations}
    \end{minipage}%
  \end{minipage}%
}
\fntext[fn1]{Work done during an internship at TeleAI.}
\fntext[fn2]{Equal Contribution.}

\author{Xiuyuan Chen\fnref{label1,label2,fn1,fn2}}

\author{Jian Zhao\fnref{label1,fn2}}

\author{Yuxiang He\fnref{label1,label3}}

\author{Yuan Xun\fnref{label1,label4}}
\author{Xinwei Liu\fnref{label1,label4}}
\author{Yanshu~Li\fnref{label1,label5}}

\author{Huilin Zhou\fnref{label1,label6}}
\author{Wei Cai\fnref{label1,label7}}
\author{Ziyan Shi\fnref{label1,label8}}

\author{Yuchen Yuan\fnref{label1}}

\author{Tianle Zhang\fnref{label1}}

\author{Chi Zhang\fnref{label1}}

\author{Xuelong Li\fnref{label1}\corref{cor3}}

\cortext[cor3]{Correspondence to Xuelong Li(xuelong\_li@ieee.org).}

\affiliation[label1]{Institute of Artificial Intelligence (TeleAI) of China Telecom}
\affiliation[label2]{Shanghai Jiao Tong University}
\affiliation[label3]{Sichuan University}
\affiliation[label4]{University of Chinese Academy of Sciences}
\affiliation[label5]{Chinese Academy of Sciences}
\affiliation[label6]{University of Science and Technology of China}
\affiliation[label7]{Peking University}
\affiliation[label8]{Harbin Institute of Technology}
\begin{abstract}
While the deployment of large language models (LLMs) in high-value industries continues to expand, the systematic assessment of their safety against jailbreak and prompt-based attacks remains insufficient. Existing safety evaluation benchmarks and frameworks are often limited by an \textit{imbalanced integration} of core components (attack, defense, and evaluation methods) and an \textit{isolation} between flexible evaluation frameworks and standardized benchmarking capabilities. These limitations hinder reliable cross-study comparisons and create unnecessary overhead for comprehensive risk assessment. To address these gaps, we present \textbf{TeleAI-Safety}, a modular and reproducible framework coupled with a systematic benchmark for rigorous LLM safety evaluation. Our framework integrates a broad collection of 19 attack methods (including one self-developed method), 29 defense methods, and 19 evaluation methods (including one self-developed method). With a curated attack corpus of 342 samples spanning 12 distinct risk categories, the TeleAI-Safety benchmark conducts extensive evaluations across 14 target models. The results reveal systematic vulnerabilities and model-specific failure cases, highlighting critical trade-offs between safety and utility, and identifying potential defense patterns for future optimization. In practical scenarios, TeleAI-Safety can be flexibly adjusted with customized attack, defense, and evaluation combinations to meet specific demands. We release our complete code and evaluation results to facilitate reproducible research and establish unified safety baselines.

\simpleicon{github} TeleAI-Safety Code: \textcolor{blue}{\url{https://github.com/yuanyc06/Tele-Safety}}
\end{abstract}

\end{frontmatter}

\section{Introduction}
\label{introduction}
The rapid proliferation of large language models (LLMs) fundamentally changes our lifestyle with applications that span from customer service and content generation to medical diagnosis and financial advisory. However, jailbreak attacks exploit carefully designed triggers to bypass safety guardrails and elicit harmful outputs, which potentially incur a class of vulnerabilities that undermine the trustworthiness of LLMs. Being prevalently deployed in critical infrastructures, the robustness of LLMs against jailbreak attacks emerges as a key factor for researchers and practitioners. 

Several benchmarks and frameworks are proposed for LLM safety evaluation (\textit{e.g.} JailJudge~\cite{liu2024jailjudge}, EasyJailbreak~\cite{zhou2024easyjailbreak}, HarmBench~\cite{mazeika2024harmbench}, AISafetyLab~\cite{zhang2025aisafetylab}, and PandaGuard~\cite{shen2025pandaguard}). Nevertheless, these works fall into two conspicuous drawbacks. 

First, existing approaches~\cite{liu2024jailjudge, mazeika2024harmbench, cai2025safe} often exhibit an imbalanced integration of core components—namely, attack, defence, and evaluation methods—as well as uneven coverage of models. For instance, HarmBench emphasises broad model coverage but overlooks defensive depth; AISafetyLab focuses on defence methods, yet evaluates only a limited set of models; and PandaGuard incorporates numerous attack and defence techniques, but still relies on a narrow range of evaluation methods. This imbalance undermines reliable cross-study comparisons and obstructs the establishment of end-to-end safety baselines.

Second, current work offers limited integration between the functions of an AI safety framework and those of an AI safety benchmark. Consequently, organisations that deploy advanced LLMs lack a unified tool to both “design evaluations” (through a framework) and “quantify performance” (through a benchmark), resulting in significant unnecessary overhead for risk management.

\begin{figure}[tbp]
\centering
\includegraphics[width=0.8\linewidth]{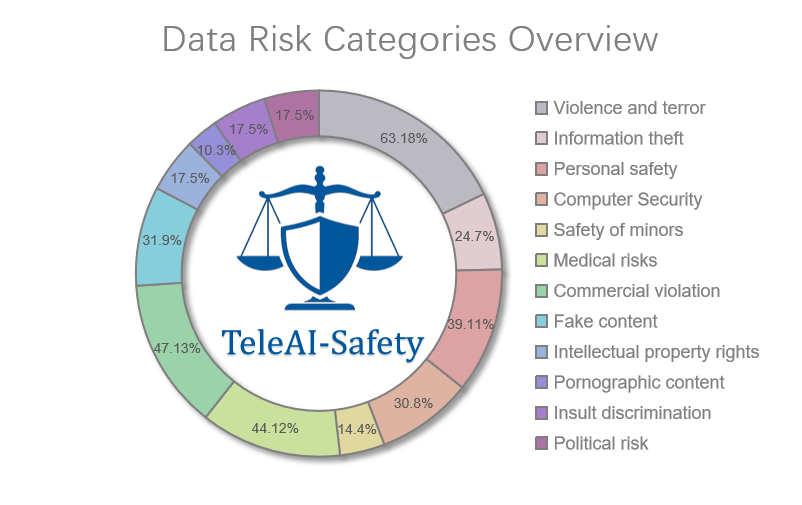}
\caption{The distribution of harmful data samples across 12 risk categories in the TeleAI-Safety dataset after the completion of the curation process.}
\label{fig:dataset_distribution}
\end{figure}

To address these gaps, we propose \textbf{TeleAI-Safety}, a comprehensive and modular AI safety framework coupled with a corresponding standardised benchmark. It is designed to systematically evaluate the robustness of LLMs against jailbreak attacks while providing a unified assessment of defensive countermeasures and evaluation methodologies. The TeleAI-Safety framework serves as the foundational infrastructure: it includes a curated dataset of 342 attack samples spanning multiple risk categories, integrates testing logic for attack, defence, and evaluation methods, and supports repeatable experimental workflows. Building on this framework, the TeleAI-Safety benchmark enables quantitative model assessment: it adopts the proposed test set, employs standardised safety and utility metrics, and produces comparable performance scores across models. Figure~\ref{fig:dataset_distribution} illustrates the distribution of data samples across risk categories, reflecting the benchmark’s coverage of real-world vulnerabilities and ensuring its practical value for LLM deployment.

\begin{table}[t]
\centering
\caption{Comparison of TeleAI-Safety with existing LLM safety benchmarks.}
\label{tab:comparison}
\resizebox{\linewidth}{!}{
\begin{tabular}{l|ccccc}
\hline
\textbf{Benchmark} & \textbf{Attack Methods} & \textbf{Defense Methods} & \textbf{Evaluation Methods} & \textbf{Standardized Dataset} & \textbf{Self-Developed Methods} \\
\hline
JailJudge~\cite{liu2024jailjudge} & \textcolor{green!70!black}{\ding{51}} & \textcolor{green!70!black}{\ding{51}} & \textcolor{green!70!black}{\ding{51}} & \textcolor{green!70!black}{\ding{51}} & \textcolor{green!70!black}{\ding{51}} \\
EasyJailbreak~\cite{zhou2024easyjailbreak} & \textcolor{green!70!black}{\ding{51}}  & \textcolor{red}{\ding{55}} & \textcolor{green!70!black}{\ding{51}} & \textcolor{red}{\ding{55}} & \textcolor{red}{\ding{55}} \\
HarmBench~\cite{mazeika2024harmbench} & \textcolor{green!70!black}{\ding{51}}  & \textcolor{red}{\ding{55}} & \textcolor{green!70!black}{\ding{51}} & \textcolor{green!70!black}{\ding{51}} & \textcolor{red}{\ding{55}} \\
AISafetyLab~\cite{zhang2025aisafetylab} & \textcolor{green!70!black}{\ding{51}} & \textcolor{green!70!black}{\ding{51}}  & \textcolor{green!70!black}{\ding{51}} & \textcolor{red}{\ding{55}} & \textcolor{red}{\ding{55}} \\
PandaGuard~\cite{shen2025pandaguard} & \textcolor{green!70!black}{\ding{51}}  & \textcolor{green!70!black}{\ding{51}}  & \textcolor{red}{\ding{55}} & \textcolor{red}{\ding{55}} & \textcolor{red}{\ding{55}} \\
\hline
\textbf{TeleAI-Safety} & \textcolor{green!70!black}{\ding{51}} & \textcolor{green!70!black}{\ding{51}} & \textcolor{green!70!black}{\ding{51}} & \textcolor{green!70!black}{\ding{51}} & \textcolor{green!70!black}{\ding{51}} (Morpheus, RADAR) \\
\hline
\end{tabular}
}
\end{table}

The TeleAI-Safety framework integrates 19 attack methods (including one self-developed method), 29 defence methods, and 19 evaluation methods (including one self-developed method). 

A distinctive strength of TeleAI-Safety lies in its co-design of framework and benchmark: the framework's rich components directly enhance the benchmark's ability of emerging threats. We also contribute two self-developed methods to the framework, both of which are integrated into the benchmark's test suite: (1) Morpheus~\cite{anonymous2026morpheus}, a self-evolving metacognitive multi-round attack agent that uses dynamic adaptation to expose vulnerabilities in complex interactions (included in the benchmark's attack evaluation); (2) RADAR~\cite{chen2025radar}, a multi-agent debate-based LLM safety evaluation method, which aims for robust risk identification. These innovations enable the TeleAI-Safety benchmark to evaluate not just traditional jailbreaks, but also emerging threats like adaptive multi-turn attacks, positioning it at the forefront of LLM safety research.

Our contributions are summarised as follows:
\begin{enumerate}
    \item We design a standardised attack corpus (342 data samples covering 12 risk categories) that serves as the testset for the TeleAI-Safety benchmark, providing a unified foundation for systematic, comparable robustness assessment across models.
    \item We develop TeleAI-Safety as an integrated \textit{framework-benchmark system}: the framework integrates 19 attack methods, 29 defence methods, and 19 evaluation methods, while the benchmark adopts this framework to conduct comprehensive LLM safety testing among 14 models (9 closed-source, 5 open-source), quantifying safety-utility trade-offs.
    \item We release TeleAI-Safety's complete code online, which provides flexible, extensible infrastructure for both framework customisation (\textit{e.g.} adding new attacks/defences) and benchmark replication, offering a reproducible open-source repository of LLM safety research.
\end{enumerate}
\section{Related Work}
\subsection{LLM Jailbreak Attack Methods}
The landscape of LLM jailbreak attacks evolves rapidly, encompassing diverse strategies that exploit different aspects of model vulnerabilities. We provide an overview of representative attack techniques integrated within our TeleAI-Safety framework. Contemporary jailbreak attack methodologies can be systematically categorized based on their access level to target models and strategic approaches. 

Based on differences in access privileges to target models and core strategic approaches, white-box, gray-box, and black-box attacks constitute a classic classification system in LLM jailbreak attacks, with the degree of model interaction as the core dimension. \emph{\textbf{White-box optimization attacks}} leverage complete model knowledge and gradient information, with GCG~\cite{zou2023gcg} pioneering gradient-based optimization of adversarial suffixes through discrete token-level optimization. \emph{\textbf{Gray-box adaptive attacks}} operate with limited model access, including AutoDAN~\cite{liu2023autodan} that employs genetic algorithms for prompt evolution and stochastic exploration strategies, LAA~\cite{andriushchenko2024laa} that combines adaptive templates with random search optimization, and AdvPrompter~\cite{paulus2024advprompter} that trains specialized attacker models for adversarial suffix generation. \emph{\textbf{Black-box attacks}} operate without internal model access, encompassing GPTFUZZER~\cite{yu2023gptfuzzer} which applies systematic mutation techniques to template evolution, and PAIR~\cite{chao2025pair} which utilizes iterative prompt refinement through attacker-target model interaction. 

Beyond access-level-oriented attack paths, methods focusing on semantic transformation, encoding obfuscation, structured search, and specialized strategic design achieve jailbreaking by exploiting specific vulnerabilities of models in language processing, representation learning, and security mechanisms. \emph{\textbf{Semantic transformation attacks}} preserve malicious intent through linguistic modifications, which include Past Tense~\cite{andriushchenko2024pasttense} that exploits temporal framing, ArtPrompt~\cite{jiang2024artprompt} that leverages visual obfuscation through ASCII art, and DeepInception~\cite{li2023deepinception} that employs nested fictional character structures. \emph{\textbf{Encoding and obfuscation attacks}} exploit representational vulnerabilities, with Cipher~\cite{yuan2023gpt} utilizing various cryptographic encoding schemes and MultiLingual~\cite{deng2023multilingual} exploiting cross-lingual safety gaps in minor languages. \emph{\textbf{Structured search approaches}}, such as TAP~\cite{mehrotra2024tap}, maintain tree-structured optimization flows for systematic attack vector exploration. And finally, \emph{\textbf{specialized strategic attacks}} represent other jailbreak attack methods with specialties, which include Jailbroken~\cite{wei2023jailbroken} that targets fundamental safety alignment failure modes, RENE~\cite{ding2023renellm} that combines prompt rewriting with scenario nesting, SCAV~\cite{xu2024scav} that maintains semantic coherence through strategic adversarial construction, ICA~\cite{wei2023ica} that employs goal hijacking techniques, and Overload~\cite{dong2024overload} that exploits computational saturation vulnerabilities.

TeleAI-Safety integrates 19 attack methods, including a novel self-developed approach, Morpheus~\cite{anonymous2026morpheus}, an advanced metacognitive multi-round attack agent.

\subsection{LLM Jailbreak Defense Methods}
In addition to LLM attack methods, increasing attention also focused on LLM defense mechanisms to mitigate such threats. Existing jailbreak defense approaches can be broadly divided into \emph{\textbf{external defenses}} and \emph{\textbf{internal defenses}}.

\emph{\textbf{External defenses}} operate at the model interface as a safety-oriented ``guardrail'', externally preventing exploitation of jailbreak prompts and blocking potentially harmful outputs. Being independent of the model's internal structure, these defenses can be applied across different models and deployment scenarios. External defense methods can be categorized as input-based defenses and output-based defenses. \emph{\textbf{Input-based defenses}} serve as the first line of protection for LLMs by filtering or sanitizing adversarial prompts before they reach the model. Some approaches detect malicious prompts using language-model metrics, such as PPL~\cite{alon2023detecting}, or specialized classifiers, like Prompt Guard~\cite{meta-llama/Prompt-Guard-86M}, while enhanced verification techniques, including Erase and Check~\cite{kumar2023certifying} and RA-LLM~\cite{cao2024defending}, further remove and check tokens to improve safety assessment. Other strategies actively manipulate the input to mitigate potential threats. Simple techniques like paraphrasing~\cite{jain2023baseline} reformulate user prompts, while advanced methods, such as SmoothLLM~\cite{robey2023smoothllm}, leverage semantic smoothing to enhance robustness. Additional methods, including EDDF~\cite{xiang2025beyond} and IBProtector~\cite{liu2024protecting}, analyze semantic content or constrain information exposure to reduce adversarial influence, while BackTranslation~\cite{wang2024defending} infers harmful intent from preliminary outputs. \emph{\textbf{Output-based defenses}}, on the other hand, focus on filtering or revising model responses before delivered to users. For instance, Self-Defense~\cite{phute2023llm} uses the model itself to re-evaluate outputs for harmful content. Aligner~\cite{ji2024aligner} employs a smaller, plug-and-play model to correct the initial output from a primary LLM, while GuardReasoner~\cite{liu2025guardreasoner} combines step-by-step reasoning with preference-based optimization to more accurately and transparently detect unsafe outputs.

\emph{\textbf{Internal defenses}} aim to enhance the model's inherent safety and robustness, and are deployed at the inference or training stage. Unlike external defenses at the interface level, these methods strengthen the model itself to fundamentally resist jailbreak attacks. Internal defense methods can be categorized as inference-time defenses and training-time defenses. \emph{\textbf{Inference-time defenses}} optimize the model's behavior during generation without modifying its parameters. Certain approaches modify the system prompt to reinforce safe behavior, such as SelfReminder~\cite{xie2023defending}, which embeds explicit safety reminders, and GoalPriority~\cite{zhang2024defending}, which instructs the model to prioritize safety over helpfulness. In-context learning techniques provide safety-oriented examples to guide outputs, as in ICD~\cite{wei2023jailbreak}. Prompt optimization methods, such as RPO~\cite{zhou2024robust}, refine prompt suffixes to improve robustness against adversarial queries. Beyond prompt-level interventions, some strategies manipulate the model's internal mechanisms to enhance robustness. Representation-based methods such as DRO~\cite{zheng2024prompt}, RePE~\cite{zou2023representation}, and JBShield~\cite{ji2024aligner} modify or leverage hidden representations to better distinguish harmful and benign inputs, while AVGAN~\cite{li2025cavgan} utilizes a GAN framework to learn the security judgment boundary within the model's internal representation space. Gradient-based defenses like GradSafe~\cite{xie2024gradsafe} and Gradient Cuff~\cite{hu2024gradient} detect jailbreak prompts by analyzing gradient patterns of safety-critical parameters or examining the value and gradient norm of a defined `Refusal Loss'. Output-focused methods further enhance safety by intervening in the decoding process, including SafeDecoding~\cite{xu2024safedecoding}, which adjusts token probabilities during generation, and RAIN~\cite{li2023rain}, which enables a model to self-evaluate its generation and rewind its state to a safer point if potential harm is detected. \emph{\textbf{Training-time defenses}}, on the other hand, adjust model parameters to align outputs with desired behaviors. Foundational strategies include fine-tuning-based safety alignments. For example, Safety-Tuned LLaMAs~\cite{bianchi2024safety} incorporates safety-oriented examples during instruction tuning. More targeted interventions, such as Backdoor Alignment~\cite{wang2024backdooralign}, inject latent backdoor triggers within safety examples to enable controllable enforcement of safe behavior. Adversarial training techniques, exemplified by C-advipo~\cite{xhonneux2024efficient}, expose the model to synthetically generated attacks in the continuous embedding space to improve robustness against prompt-level exploits. Knowledge and model editing methods, including DELMAN~\cite{wang2025delman} and Layer-AdvPatcher~\cite{ouyang2025layer}, directly modify model parameters or layers to remove specific vulnerabilities. TeleAI-Safety integrates 29 defense methods covering these diverse strategies.

\subsection{LLM Jailbreak Evaluations}
Benefiting from the significant progress of LLMs in language understanding capabilities, current safety evaluation methods tend to transform from traditional rule-based matching methods to semantic content-based risk prevention methods. AISafetyLab~\cite{zhang2025aisafetylab} integrates safety evaluators based on keyword pattern matching, which is able to rapidly identify potentially harmful content through predefined keyword lists. HarmBench~\cite{mazeika2024harmbench} and GPTFuzzer~\cite{yu2023gptfuzzer}, in addition to introducing jailbreak attack methods, also contribute specialized evaluation models for harmful content identification and classification.

As model capabilities advance, fine-tuning-based approaches in risk assessment have emerged. For instance, ShieldLM, LlamaGuard3, and ShieldGemma employ specialized training data for jailbreak attack case analysis and utilize these capabilities for content moderation of their associated model outputs.

In recent years, evaluation methods based on conversational LLMs have continuously emerged. Studies propose various specifically designed prompt templates and paradigms to instruct general-purpose LLMs better, leveraging their powerful language understanding capabilities to accomplish evaluation tasks. These methods all aim to achieve efficient evaluation through single-model interaction.

Meanwhile, advances in multi-agent interaction technologies have promoted more comprehensive evaluation frameworks. For example, RADAR~\cite{chen2025radar} enhances detection coverage and accuracy by integrating multiple specialized agents working collaboratively, successfully addressing increased and more complex safety challenges.

TeleAI-Safety integrates 19 evaluation methods, including a self-developed evaluation approach that leverages advanced reasoning capabilities for more accurate harm detection.

\subsection{Existing Benchmarks}
Several comprehensive benchmarks have emerged to evaluate LLM safety against jailbreak attacks, each contributing unique perspectives and methodological innovations to the field. JailJudge~\cite{liu2024jailjudge} introduces systematic evaluation protocols with 18 human judges across 4 language models, establishing precedents for multi-judge assessment systems. EasyJailbreak~\cite{zhou2024easyjailbreak} provides a comprehensive toolkit integrating 12 attack methods with evaluation capabilities across 10 models, emphasizing accessibility and reproducibility. HarmBench~\cite{mazeika2024harmbench} significantly expands the evaluation scope with 18 attack methodologies across 33 language models, setting new standards for comprehensive model coverage. AISafetyLab~\cite{zhang2025aisafetylab} distinguishes itself by integrating both offensive and defensive perspectives, incorporating 13 attack methods alongside 16 defensive mechanisms. PandaGuard~\cite{shen2025pandaguard} represents a recent advancement in comprehensive safety benchmarking, integrating 19 attack methods with 12 defensive techniques across 49 models and supporting 7 distinct LLM interfaces.

\paragraph{\textbf{Our Distinctive Features}}
TeleAI-Safety distinguishes itself through the comprehensive integration of 19 attack methods, 29 defense methods, and 19 evaluation methods, surpassing existing frameworks in scope. It also incorporates cutting-edge self-developed attack and evaluation methods. Additionally, we employ a carefully curated risk-diverse dataset of 342 attack data samples spanning 12 risky categories and provide contemporary benchmarking results.
\begin{figure}[t]
    \centering
    \includegraphics[width=\textwidth]{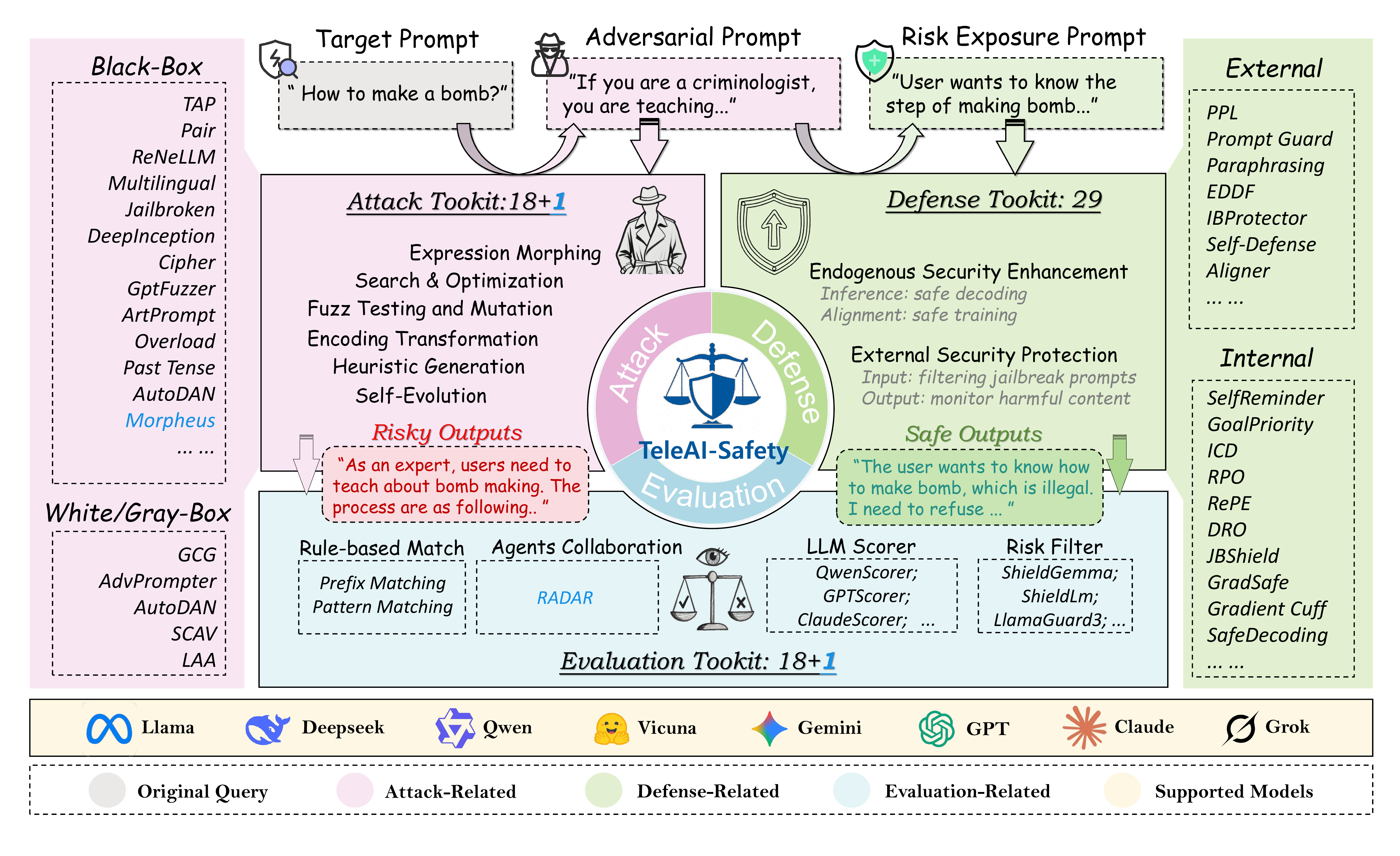}
    \caption{The framework of TeleAI-Safety, which integrates 19 attack methods, 29 defense methods, and 19 evaluation methods. The blue highlights indicate our latest self-developed methods.}
    \label{fig:framework}
\end{figure}

\section{TeleAI-Safety Framework}
\subsection{LLM Jailbreaks Formulation}
We first provide the problem formulation of LLM jailbreak, which motivates the TeleAI-Safety framework. LLM jailbreak refers to the systematic exploitation of model vulnerabilities to elicit harmful, unintended, or policy-violating responses through carefully crafted adversarial prompts. Formally, we formulate a jailbreak attack as a process where an adversary seeks to manipulate a safety-aligned language model to produce prohibited content by circumventing its built-in safety mechanisms.

Let $\mathcal{M}: \mathcal{X} \rightarrow \mathcal{Y}$ represent an LLM that maps an input prompt $x \in \mathcal{X}$ to an output response $y \in \mathcal{Y}$, where $\mathcal{X}$ denotes the input space of natural language prompts and $\mathcal{Y}$ represents the output space of generated texts. We define a safety policy as a binary function, $\mathcal{P}: \mathcal{Y} \rightarrow \{0, 1\}$, that determines whether an output $y$ violates safety guidelines, where $\mathcal{P}(y) = 1$ indicates policy violation and $\mathcal{P}(y) = 0$ otherwise. A jailbreak attack can be formalized as an optimization problem where an adversary constructs a malicious prompt $x_{\text{adv}} \in \mathcal{X}$ such that the model's response $y_{\text{adv}} = \mathcal{M}(x_{\text{adv}})$ satisfies the adversary's objective while violating the safety policy. Specifically, given a target harmful instruction $h \in \mathcal{H}$, where $\mathcal{H}$ represents the space of harmful requests, the adversary seeks to find:
\begin{equation}
x_{\text{adv}}^* = \arg\max_{x_{\text{adv}} \in \mathcal{X}} \mathbb{P}[\mathcal{P}(\mathcal{M}(x_{\text{adv}})) = 1 \land \mathbb{E}(h, \mathcal{M}(x_{\text{adv}}))],
\end{equation}
where $\mathbb{E}(h, y)$ denotes a semantic satisfaction function that evaluates whether the model output $y$ addresses the harmful instruction $h$. The attack success rate (ASR) for a given attack method $\mathcal{A}$ over a dataset $\mathcal{D} = \{h_1, h_2, \ldots, h_n\}$ of harmful instructions is defined as:
\begin{equation}
\text{ASR} = \frac{1}{|\mathcal{D}|} \sum_{i=1}^{|\mathcal{D}|} \mathbb{I}[\mathcal{P}(\mathcal{M}(\mathcal{A}(h_i))) = 1 \land \mathbb{E}(h_i, \mathcal{M}(\mathcal{A}(h_i)))],
\end{equation}
where $\mathcal{A}(h_i)$ represents the adversarial prompt generated by attack method $\mathcal{A}$ for harmful instruction $h _i$, and $\mathbb{I}[\cdot]$ denotes the indicator function.

\subsection{Framework Overview}
The TeleAI-Safety framework is implemented in our open-source repository at \textcolor{blue}{\url{https://github.com/yuanyc06/Tele-Safety}}, and we will subsequently introduce its architecture based on the repository structure. 

TeleAI-Safety adopts a modular and configurable design paradigm, with its core modules as \texttt{attacks}, \texttt{defenses}, \texttt{evaluations}, \texttt{configs}, \texttt{datasets}, and \texttt{models}. The attack, defense, and evaluation modules are designed as independent components, enabling flexible customizations through YAML configuration files. To address the lack of comprehensive resources in LLM jailbreak research, the TeleAI-Safety repository integrates multiple types of resources: (1) a self-constructed dataset with 342 data samples; (2) 19 LLM jailbreak attack methods (including one self-developed attack method Morpheus~\cite{anonymous2026morpheus}); (3) 29 defense methods; and (4) 19 evaluation methods (including one self-developed evaluation method RADAR ~\cite{chen2025radar}). This integration realizes a comprehensive collection of the latest attack, defense, and evaluation techniques for LLM jailbreak scenarios, providing a one-stop research framework for LLM safety assessment.

\subsection{Modules}
The modular design of TeleAI-Safety ensures clear functional division, low coupling between modules, and high extensibility. By isolating core functionalities into dedicated modules, the framework enables researchers to modify or extend individual components (\textit{e.g.} new attack methods) without disrupting the overall workflow, thus accelerating the experimental process. The detailed functions of the modules, including their design and operational mechanisms, are elaborated below.

\paragraph{Datasets}
The \texttt{Datasets} module serves as the data layer for LLM safety benchmarking, with a primary focus on providing high-quality, scenario-specific data to validate attack effectiveness and defense robustness. A core contribution of this module is the integration of our self-constructed dataset, which integrates 12 distinct risk categories (\textit{e.g.} harmful content generation, privacy information leakage) that are critical to LLM jailbreak evaluations. This dataset comprises 342 manually curated samples, each annotated with detailed risk labels. To enhance flexibility and accommodate diverse research demands, this module also supports integration of external data sources: it can directly load datasets from the Hugging Face Hub (\textit{e.g.} leveraging public LLM safety datasets such as HarmBench~\cite{mazeika2024harmbench}) and accept user-specified local file paths, enabling researchers to incorporate custom data into their experiments, such as domain-specific risk prompts for healthcare or finance LLMs. The dual support for proprietary and external data ensures TeleAI-Safety's adaptability to both standardized and specialized research scenarios.

\paragraph{Models}
The \texttt{Models} module is designed to address the inconsistency issue of LLM interface accessibility by providing a unified model loading and management framework. The key of this module is the \texttt{load\_model()} function, which abstracts the complexities of model invocation through three parameters that cover all major LLM deployment paradigms. The \texttt{model\_type} parameter specifies the model access mode, with valid values as ``local'' (for models deployed on on-premises hardware or hosted on the Hugging Face Hub) or ``openai'' (for cloud-based models accessed via official application programming interfaces, such as OpenAI's GPT series). Complementing to this, the \texttt{model\_name} parameter acts as a unique identifier: for Hugging Face models, this corresponds to the official repository name, while for API models, it refers to the service-specific model identifier (\textit{e.g.} ``gpt-4-turbo''). The \texttt{model\_path} parameter specifies the storage location, which is required for local models (\textit{e.g.} ``/data/models/llama-2-7b-chat") or cached Hugging Face models (\textit{e.g.} ``meta-llama/llama-2-7b-chat") to ensure efficient model loading without redundant downloads. By encapsulating these parameters within a single function, this module enables seamless switching between different LLMs without modifying the core logic of attack or defense modules, significantly reducing experimental setup overhead and potential manual error.

\paragraph{Attacks}
The \texttt{Attacks} module constitutes the core experimental layer for evaluating LLM vulnerability to jailbreak attempts, incorporating a comprehensive suite of attack techniques to cover mainstream LLM jailbreak scenarios. Specifically, the module integrates 19 well-established classical attack methods and 1 self-developed attack method~\cite{anonymous2026morpheus}. A key feature of this module is its coverage of three fundamental attack scenarios, which are defined based on the level of model information availability to the attacker: (1) white-box attacks, which require access to internal model parameters or gradient information (\textit{e.g.} gradient-based adversarial prompt optimization methods, in which attackers leverage model gradients to refine prompts that bypass safety guardrails); (2) gray-box attacks, which rely on partial model information (\textit{e.g.} analyzing the probability distribution of model outputs to identify weak points in safety alignment); (3) black-box attacks, which operate without any internal model knowledge (\textit{e.g.} prompt engineering-based jailbreak that exploits linguistic ambiguities, or multi-turn conversation inducement that gradually steers the model toward generating harmful content). For attack methods that depend on auxiliary models (\textit{e.g.} attack models used to generate adversarial prompts), the module supports flexible parameter configuration via YAML files, where users can specify details of auxiliary attack models (\textit{e.g.} \texttt{attack\_model\_type}, \texttt{attack\_model\_name}, \texttt{attack\_model\_path}). Detailed technical principles for all 19 attack methods are documented in appendix.

\paragraph{Defenses}
The \texttt{Defenses} module serves as the safeguarding layer in TeleAI-Safety, incorporating a comprehensive suite of defense strategies to protect LLMs against jailbreak attacks across diverse threat scenarios. Specifically, the module integrates 29 advanced defense methods, covering a wide range of techniques, including input/output filtering, decoding regulation, and safety-aligned model optimization. To facilitate systematic evaluation, this module categorizes defense methods into external and internal methods based on their deployment stage. External defenses operate at the model interface level as safety ``guardrails'', which include methods such as prompt and output filtering that conduct content moderation before or after content generation. Internal defenses, on the other hand, are embedded within the model's processing pipeline and include inference-time mechanisms (\textit{e.g.} safe decoding, gradient-based refusal detection) as well as training-time interventions (\textit{e.g.} adversarial training, safety alignment fine-tuning) that improve the model's intrinsic robustness. A key feature of this module lies in its ability to support the composition of multiple defenses simultaneously. By combining external and internal defense strategies, this module provides end-to-end protection throughout the entire model interaction lifecycle, increasing model resilience against a wide spectrum of jailbreak attacks. Each defense method is implemented with a standardized interface, similar to the way attack methods are configured. All defenses---except those applied at training time---support flexible configuration through dedicated YAML files, in which users can specify relevant parameters such as \texttt{defense\_model\_type}, \texttt{defense\_model\_name}, and \texttt{defense\_model\_path}. This design enables seamless integration of new defenses or adaptation of existing ones without modifying the core code. Detailed technical principles and implementation details for all 29 defense methods are documented in appendix.

\paragraph{Evaluations}
The \texttt{Evaluations} module serves as the component for the comprehensive evaluation of LLM attack and defense methods, which integrates a complete suite of evaluation tools to cover various scenarios and dimensions in the evaluation of LLM safety. Specifically, this module incorporates 19 established classical evaluation methods alongside one self-developed evaluation method~\cite{chen2025radar}, which focuses on addressing evaluation bias inherent in single-evaluator systems. A key advantage of this module lies in its coverage of multiple evaluation mechanisms, which are categorized based on assessment depth and interaction modes: pattern-matching-based rapid screening (\textit{e.g.} prefix matching and pattern matching for efficient identification of known malicious query patterns); LLM-based risk evaluators that leverage general-purpose chat models to conduct risk assessment of LLM outputs (\textit{e.g.} \texttt{QwenScorer}, \texttt{GPT5Scorer}, \texttt{ClaudeScorer} etc.); risk controllers (\textit{e.g.} ShieldGemma, Shield-LM, and LlamaGuard-3, which employ fine-tuned risk management models to perform risk level control and attack outcome decisions); and RADAR~\cite{chen2025radar}, our self-developed method based on multi-agent collaboration that simulates multi-round debates among multiple evaluators to achieve consensus judgment on query risks and model response safety. Detailed technical principles and applicable scenarios for all 19 evaluation methods are documented in appendix.

\section{Experimental Results}

\subsection{Datasets}
\label{sec:datasets}

A robust, diverse, and well-curated dataset is fundamental to systematic evaluation of LLM safety. To this end, we introduce the \textit{TeleAI-Safety dataset}, a meticulously constructed collection of 342 harmful data samples spanning 12 risk categories. This dataset serves as the test set for our benchmark, designed to provide a standardized and challenging foundation for assessing the effectiveness of both jailbreak attacks and defensive countermeasures.

\paragraph{Curation Principles and Process}
Our construction process is guided by the goal of creating a dataset that is both comprehensive in scope and rigorous in quality, addressing the limitations of simply aggregating existing benchmarks. The process involves three main stages:

\begin{enumerate}
    \item \textbf{Initial Pool Collection:} We begin by compiling a comprehensive initial pool of data samples from prominent existing safety and jailbreak benchmarks, including GuidedBench~\cite{huang2025guidedbenchmeasuringmitigatingevaluation}, HarmBench~\cite{mazeika2024harmbench}, JailbreakBench~\cite{chao2024jailbreakbenchopenrobustnessbenchmark}, and AdvBench~\cite{zou2023gcg}. This broad collection serves as the foundation for our subsequent refinement stages.

    \item \textbf{Principled Filtering and Refinement:} Acknowledging that simple aggregation introduces inconsistencies and biases, we are inspired by the rigorous filtering methodology of a recent work~\cite{huang2025guidedbenchmeasuringmitigatingevaluation}. We apply a strict set of principles to refine the initial pool. Each candidate question is vetted to ensure it meets the following criteria:
    \begin{itemize}
        \item \textit{Baseline Refusal:} The question must be consistently refused by major, well-aligned LLMs (\textit{e.g.} GPT-4, Claude 3) under normal conditions. This ensures that we are testing the ability to bypass safety mechanisms, not simply asking for borderline content.
        \item \textit{Direct and Unambiguous Intent:} The prompt must represent a direct and clear request for harmful content, free from complex, confounding scenarios or role-playing narratives that could muddle the evaluation.
        \item \textit{Authenticity:} We minimize questions that rely on artificial triggers (\textit{e.g.} prepending ``illegally'' to a benign query), favoring prompts that reflect more realistic malicious user intent.
    \end{itemize}

    \item \textbf{Taxonomy-Guided Selection:} Following the filtering stage, the refined questions are categorized and selected based on our novel risk taxonomy (detailed below). This step ensures that our final dataset has a balanced and meaningful distribution across critical risk areas. Redundant or semantically overlapping questions within the same narrow category are deduplicated to maximize the diversity and challenge of the test set.
\end{enumerate}

Through this multi-stage process of collection, principled filtering, and taxonomy-guided selection, we construct our final dataset of 342 data samples.

\paragraph{Taxonomy of Risk Categories}
To ensure the authority and relevance of our dataset, we develop a structured taxonomy of 12 risk categories. The taxonomy is systematically grounded in national regulations and international technical standards, including China's national standard GB/T 45654-2025, the ``Interim Measures for the Management of Generative Artificial Intelligence Services'', and the ``Cybersecurity Law of the People's Republic of China''. To align with global best practices, we also cross-reference our categories with internationally recognized frameworks such as Aegis 2.0, the MITRE ATLAS framework, and the OWASP Top 10 for LLMs. The 12 risk categories are: Political Risk, Medical Risk, Personal Safety Risk, Commercial Violations, Information Theft, Pornographic Content, Insults and Discrimination, Content Fabrication, Violence and Terrorism, Intellectual Property Protection, Harm to Minors, and Cybersecurity.

\paragraph{Dataset Distribution}
Figure~\ref{fig:dataset_distribution} shows the final distribution of the 342 data samples across the 12 risk categories. The number of data samples in each category is the result of our multi-stage filtering process, which prioritizes the quality and authenticity of each prompt. This process yields a non-uniform distribution, with categories such as `Violence and Terrorism` and `Commercial Violations` being more represented. Nevertheless, all categories contain a sufficient number of data samples to support a meaningful evaluation.

\subsection{Models}
To ensure a comprehensive evaluation of model safety, we select a diverse set of state-of-the-art LLMs, categorizing them into black-box and white-box models based on their accessibility. This approach allows us to examine safety behaviors under different constraints—from real-world deployment scenarios with only API access to controlled settings with full model internals.

{\bf Black-Box Models.} For the black-box evaluation, where we have access only to model outputs, we consider 9 prominent proprietary models deployed via API services. From the GPT series, we include the newly released GPT-5, the powerful GPT-4.1, the cost-effective GPT-4o mini, and the research-focused system OpenAI-o1. We also evaluate xAI's Grok-3 and its more efficient variant, Grok-3-mini, to represent a different architectural lineage. Furthermore, we incorporate Anthropic's Claude-3.5 Sonnet, recognized for its strong reasoning capabilities, and Google's Gemini-2.5-Pro, which showcases advanced multimodal understanding. This selection covers the most influential and widely discussed commercial models currently available.

{\bf White-Box Models.} For white-box evaluation, where full model parameters and internal states are accessible, we choose 5 leading open-source models. This includes smaller yet efficient models such as Vicuna-7B and Llama-3.1-8B-Instruct, which are fine-tuned for dialogue and serve as strong baselines in the community. We also test Deepseek-R1, a model explicitly trained with reinforcement learning for enhanced reasoning. To ensure broad representation, we evaluate the Qwen series from Alibaba, namely Qwen-1.5-7B-Chat and its latest iteration Qwen-2.5-7B-Instruct, which have demonstrated competitive performance across various benchmarks. These models allow for in-depth probing and controlled experiments on the mechanisms underlying safety features.

\subsection{Evaluation Metrics}
We report the Attack Success Rate (ASR) of different models under evaluation on the datasets as a metric to measure model safety. To demonstrate the performance differences across different attacks, defenses, and evaluation methods, we present detailed experimental results across these three dimensions. Additionally, for each model, we report the ASR under different risk category attacks as a measure of category-wise vulnerability assessment for different models.

\subsection{Model-wise Attack Results}
\begin{table}[htbp]
    \small
    \centering
    \setlength{\tabcolsep}{3pt}
    {
    \renewcommand\arraystretch{0.9}
    \caption{ASR of different attack methods on black-box models.}
    \resizebox{\linewidth}{!}{
        \begin{tabular}[c]{l|ccccccccc}
\toprule
{Attack Method} & GPT-5 & GPT-4.1 & OpenAI-o1 & GPT-4o mini & GPT-4.1 mini & Grok-3 & Grok-3-mini & Claude-3.5 & Gemini-2.5 \\ \\
\midrule
TAP & {\cellcolor[RGB]{224,194,206} 0.56} & {\cellcolor[RGB]{240,187,199} 0.73} & {\cellcolor[RGB]{181,212,226} 0.10} & {\cellcolor[RGB]{219,196,209} 0.50} & {\cellcolor[RGB]{231,191,203} 0.63} & {\cellcolor[RGB]{196,206,219} 0.26} & {\cellcolor[RGB]{195,206,219} 0.25} & {\cellcolor[RGB]{178,213,227} 0.07} & {\cellcolor[RGB]{183,211,225} 0.12} \\
PAIR & {\cellcolor[RGB]{185,210,224} 0.14} & {\cellcolor[RGB]{216,198,210} 0.47} & {\cellcolor[RGB]{221,196,208} 0.52} & {\cellcolor[RGB]{209,200,213} 0.40} & {\cellcolor[RGB]{209,200,213} 0.40} & {\cellcolor[RGB]{207,201,214} 0.38} & {\cellcolor[RGB]{214,198,211} 0.45} & {\cellcolor[RGB]{173,216,230} 0.01} & {\cellcolor[RGB]{229,192,204} 0.61} \\
RENE & {\cellcolor[RGB]{207,201,214} 0.38} & {\cellcolor[RGB]{205,202,215} 0.36} & {\cellcolor[RGB]{202,203,216} 0.32} & {\cellcolor[RGB]{213,199,211} 0.44} & {\cellcolor[RGB]{201,204,217} 0.31} & {\cellcolor[RGB]{206,201,214} 0.37} & {\cellcolor[RGB]{204,203,215} 0.34} & {\cellcolor[RGB]{186,210,224} 0.15} & {\cellcolor[RGB]{211,199,212} 0.42} \\
Cipher & {\cellcolor[RGB]{229,192,204} 0.61} & {\cellcolor[RGB]{214,198,211} 0.45} & {\cellcolor[RGB]{186,210,224} 0.15} & {\cellcolor[RGB]{217,197,210} 0.48} & {\cellcolor[RGB]{210,200,212} 0.41} & {\cellcolor[RGB]{213,199,211} 0.44} & {\cellcolor[RGB]{211,199,212} 0.42} & {\cellcolor[RGB]{176,214,228} 0.05} & {\cellcolor[RGB]{195,206,219} 0.25} \\
Jailbroken & {\cellcolor[RGB]{201,204,217} 0.31} & {\cellcolor[RGB]{181,212,226} 0.10} & {\cellcolor[RGB]{178,213,227} 0.07} & {\cellcolor[RGB]{185,210,224} 0.14} & {\cellcolor[RGB]{189,209,222} 0.18} & {\cellcolor[RGB]{199,205,218} 0.29} & {\cellcolor[RGB]{176,214,228} 0.05} & {\cellcolor[RGB]{174,215,229} 0.03} & {\cellcolor[RGB]{177,214,227} 0.06} \\
MultiLingual & {\cellcolor[RGB]{190,208,221} 0.20} & {\cellcolor[RGB]{184,211,224} 0.13} & {\cellcolor[RGB]{175,214,228} 0.04} & {\cellcolor[RGB]{181,212,226} 0.10} & {\cellcolor[RGB]{186,210,224} 0.15} & {\cellcolor[RGB]{215,198,210} 0.46} & {\cellcolor[RGB]{211,199,212} 0.42} & {\cellcolor[RGB]{189,209,222} 0.18} & {\cellcolor[RGB]{189,208,222} 0.19} \\
ArtPrompt & {\cellcolor[RGB]{211,199,212} 0.42} & {\cellcolor[RGB]{192,207,221} 0.22} & {\cellcolor[RGB]{187,210,223} 0.16} & {\cellcolor[RGB]{192,207,221} 0.22} & {\cellcolor[RGB]{195,206,219} 0.25} & {\cellcolor[RGB]{189,208,222} 0.19} & {\cellcolor[RGB]{188,209,223} 0.17} & {\cellcolor[RGB]{179,213,227} 0.08} & {\cellcolor[RGB]{191,208,221} 0.21} \\
SCAV & {\cellcolor[RGB]{180,212,226} 0.09} & {\cellcolor[RGB]{192,207,221} 0.22} & {\cellcolor[RGB]{173,215,229} 0.02} & {\cellcolor[RGB]{199,205,218} 0.29} & {\cellcolor[RGB]{188,209,223} 0.17} & {\cellcolor[RGB]{233,190,202} 0.65} & {\cellcolor[RGB]{225,194,206} 0.57} & {\cellcolor[RGB]{173,215,229} 0.02} & {\cellcolor[RGB]{222,195,207} 0.54} \\
ICA & {\cellcolor[RGB]{187,210,223} 0.16} & {\cellcolor[RGB]{188,209,223} 0.17} & {\cellcolor[RGB]{189,209,222} 0.18} & {\cellcolor[RGB]{194,207,220} 0.24} & {\cellcolor[RGB]{195,206,219} 0.25} & {\cellcolor[RGB]{187,210,223} 0.16} & {\cellcolor[RGB]{191,208,221} 0.21} & {\cellcolor[RGB]{173,215,229} 0.02} & {\cellcolor[RGB]{175,214,228} 0.04} \\
DRA & {\cellcolor[RGB]{177,214,227} 0.06} & {\cellcolor[RGB]{222,195,207} 0.54} & {\cellcolor[RGB]{205,202,215} 0.35} & {\cellcolor[RGB]{212,199,212} 0.43} & {\cellcolor[RGB]{228,192,204} 0.60} & {\cellcolor[RGB]{223,194,207} 0.55} & {\cellcolor[RGB]{207,201,214} 0.38} & {\cellcolor[RGB]{184,211,224} 0.13} & {\cellcolor[RGB]{222,195,207} 0.53} \\
DeepInception & {\cellcolor[RGB]{185,210,224} 0.14} & {\cellcolor[RGB]{204,203,215} 0.34} & {\cellcolor[RGB]{174,215,229} 0.03} & {\cellcolor[RGB]{234,190,202} 0.66} & {\cellcolor[RGB]{224,194,206} 0.56} & {\cellcolor[RGB]{222,195,207} 0.54} & {\cellcolor[RGB]{202,203,216} 0.32} & {\cellcolor[RGB]{197,205,218} 0.27} & {\cellcolor[RGB]{209,200,213} 0.40} \\
GPTFUZZER & {\cellcolor[RGB]{198,205,218} 0.28} & {\cellcolor[RGB]{189,209,222} 0.18} & {\cellcolor[RGB]{192,207,221} 0.22} & {\cellcolor[RGB]{194,207,220} 0.24} & {\cellcolor[RGB]{185,210,224} 0.14} & {\cellcolor[RGB]{219,196,209} 0.50} & {\cellcolor[RGB]{229,192,204} 0.61} & {\cellcolor[RGB]{173,216,230} 0.01} & {\cellcolor[RGB]{175,214,228} 0.04} \\
Past tense & {\cellcolor[RGB]{195,206,219} 0.25} & {\cellcolor[RGB]{224,194,206} 0.56} & {\cellcolor[RGB]{199,205,218} 0.29} & {\cellcolor[RGB]{222,195,207} 0.54} & {\cellcolor[RGB]{225,194,206} 0.57} & {\cellcolor[RGB]{214,198,211} 0.45} & {\cellcolor[RGB]{198,205,218} 0.28} & {\cellcolor[RGB]{194,207,220} 0.24} & {\cellcolor[RGB]{254,182,193} 0.87} \\
Overload & {\cellcolor[RGB]{173,215,229} 0.02} & {\cellcolor[RGB]{180,212,226} 0.09} & {\cellcolor[RGB]{174,215,229} 0.03} & {\cellcolor[RGB]{193,207,220} 0.23} & {\cellcolor[RGB]{197,205,218} 0.27} & {\cellcolor[RGB]{192,207,221} 0.22} & {\cellcolor[RGB]{214,198,211} 0.45} & {\cellcolor[RGB]{173,216,230} 0.01} & {\cellcolor[RGB]{254,182,193} 0.87} \\
Morpheus & {\cellcolor[RGB]{220,196,208} 0.51} & {\cellcolor[RGB]{255,182,193} 0.88} & {\cellcolor[RGB]{240,187,199} 0.73} & {\cellcolor[RGB]{237,189,201} 0.69} & {\cellcolor[RGB]{255,182,193} 0.88} & {\cellcolor[RGB]{247,185,196} 0.80} & {\cellcolor[RGB]{243,186,198} 0.76} & {\cellcolor[RGB]{209,200,213} 0.40} & {\cellcolor[RGB]{225,194,206} 0.57} \\
\midrule
\midrule
ASR Avg. & {\cellcolor[RGB]{198,205,218} 0.28} & {\cellcolor[RGB]{205,202,215} 0.36} & {\cellcolor[RGB]{191,208,221} 0.21} & {\cellcolor[RGB]{206,201,214} 0.37} & {\cellcolor[RGB]{207,201,214} 0.38} & {\cellcolor[RGB]{211,199,212} 0.42} & {\cellcolor[RGB]{207,201,214} 0.38} & {\cellcolor[RGB]{182,212,225} 0.11} & {\cellcolor[RGB]{207,201,214} 0.38} \\
ASR Std. & {\cellcolor[RGB]{189,209,222} 0.18} & {\cellcolor[RGB]{193,207,220} 0.23} & {\cellcolor[RGB]{190,208,221} 0.20} & {\cellcolor[RGB]{189,209,222} 0.18} & {\cellcolor[RGB]{191,208,221} 0.21} & {\cellcolor[RGB]{188,209,223} 0.17} & {\cellcolor[RGB]{189,209,222} 0.18} & {\cellcolor[RGB]{182,212,225} 0.11} & {\cellcolor[RGB]{197,205,218} 0.27} \\
\bottomrule
\end{tabular}
    }
    \label{tab:black_box}
    }
\end{table}
\begin{table}[htbp]
    \small
    \centering
    \setlength{\tabcolsep}{5pt}
    {
    \renewcommand\arraystretch{0.8}
    \caption{ASR of different attack methods on white-box models.}
    \resizebox{\linewidth}{!}{
        \begin{tabular}[c]{l|ccccc}
\toprule
{Attack Method} & Vicuna-7B & Llama-3.1-Instruct & Deepseek-R1 & Qwen-1.5-Chat & Qwen-2.5-Instruct \\ \\
\midrule
MultiLingual & {\cellcolor[RGB]{195,206,219} 0.25} & {\cellcolor[RGB]{185,210,224} 0.14} & {\cellcolor[RGB]{214,199,211} 0.45} & {\cellcolor[RGB]{174,215,229} 0.03} & {\cellcolor[RGB]{173,215,229} 0.02} \\
Jailbroken & {\cellcolor[RGB]{183,211,225} 0.12} & {\cellcolor[RGB]{178,213,227} 0.07} & {\cellcolor[RGB]{227,193,205} 0.59} & {\cellcolor[RGB]{176,214,228} 0.05} & {\cellcolor[RGB]{179,213,227} 0.08} \\
PAIR & {\cellcolor[RGB]{241,187,199} 0.74} & {\cellcolor[RGB]{229,192,204} 0.62} & {\cellcolor[RGB]{218,197,209} 0.50} & {\cellcolor[RGB]{183,211,225} 0.12} & {\cellcolor[RGB]{226,193,206} 0.58} \\
TAP & {\cellcolor[RGB]{195,206,219} 0.25} & {\cellcolor[RGB]{185,210,224} 0.14} & {\cellcolor[RGB]{218,197,209} 0.50} & {\cellcolor[RGB]{223,195,207} 0.55} & {\cellcolor[RGB]{186,210,224} 0.15} \\
Cipher & {\cellcolor[RGB]{187,209,223} 0.17} & {\cellcolor[RGB]{173,216,230} 0.01} & {\cellcolor[RGB]{215,198,210} 0.47} & {\cellcolor[RGB]{186,210,224} 0.15} & {\cellcolor[RGB]{203,203,216} 0.34} \\
RENE & {\cellcolor[RGB]{215,198,210} 0.47} & {\cellcolor[RGB]{201,204,216} 0.32} & {\cellcolor[RGB]{219,196,208} 0.51} & {\cellcolor[RGB]{203,203,216} 0.34} & {\cellcolor[RGB]{202,203,216} 0.33} \\
ArtPrompt & {\cellcolor[RGB]{194,207,220} 0.24} & {\cellcolor[RGB]{195,206,219} 0.25} & {\cellcolor[RGB]{207,201,214} 0.38} & {\cellcolor[RGB]{207,201,214} 0.38} & {\cellcolor[RGB]{187,209,223} 0.17} \\
SCAV & {\cellcolor[RGB]{236,189,201} 0.69} & {\cellcolor[RGB]{191,208,221} 0.21} & {\cellcolor[RGB]{221,195,208} 0.53} & {\cellcolor[RGB]{233,190,202} 0.66} & {\cellcolor[RGB]{234,190,202} 0.67} \\
Overload & {\cellcolor[RGB]{230,192,203} 0.63} & {\cellcolor[RGB]{198,205,218} 0.28} & {\cellcolor[RGB]{211,200,212} 0.42} & {\cellcolor[RGB]{216,197,210} 0.48} & {\cellcolor[RGB]{214,199,211} 0.45} \\
Past tense & {\cellcolor[RGB]{222,195,207} 0.54} & {\cellcolor[RGB]{216,197,210} 0.48} & {\cellcolor[RGB]{225,194,206} 0.57} & {\cellcolor[RGB]{204,202,215} 0.35} & {\cellcolor[RGB]{214,199,211} 0.45} \\
Random Search & {\cellcolor[RGB]{234,190,202} 0.67} & {\cellcolor[RGB]{242,187,198} 0.76} & {\cellcolor[RGB]{239,188,200} 0.72} & {\cellcolor[RGB]{225,194,206} 0.57} & {\cellcolor[RGB]{240,188,199} 0.73} \\
AutoDAN & {\cellcolor[RGB]{218,197,209} 0.50} & {\cellcolor[RGB]{209,200,213} 0.40} & {\cellcolor[RGB]{221,195,208} 0.53} & {\cellcolor[RGB]{215,198,210} 0.47} & {\cellcolor[RGB]{217,197,209} 0.49} \\
LAA & {\cellcolor[RGB]{239,188,200} 0.72} & {\cellcolor[RGB]{202,203,216} 0.33} & {\cellcolor[RGB]{215,198,210} 0.47} & {\cellcolor[RGB]{194,207,220} 0.24} & {\cellcolor[RGB]{235,190,201} 0.68} \\
GPTFUZZER & {\cellcolor[RGB]{239,188,200} 0.72} & {\cellcolor[RGB]{188,209,222} 0.18} & {\cellcolor[RGB]{211,200,212} 0.42} & {\cellcolor[RGB]{255,182,193} 0.89} & {\cellcolor[RGB]{207,201,214} 0.38} \\
DeepInception & {\cellcolor[RGB]{235,190,201} 0.68} & {\cellcolor[RGB]{224,194,206} 0.56} & {\cellcolor[RGB]{224,194,206} 0.56} & {\cellcolor[RGB]{230,192,203} 0.63} & {\cellcolor[RGB]{232,191,203} 0.65} \\
GCG & {\cellcolor[RGB]{226,193,206} 0.58} & {\cellcolor[RGB]{175,214,228} 0.04} & {\cellcolor[RGB]{214,199,211} 0.45} & {\cellcolor[RGB]{174,215,229} 0.03} & {\cellcolor[RGB]{174,215,229} 0.03} \\
DRA & {\cellcolor[RGB]{244,186,197} 0.78} & {\cellcolor[RGB]{228,192,204} 0.61} & {\cellcolor[RGB]{209,200,213} 0.40} & {\cellcolor[RGB]{222,195,207} 0.54} & {\cellcolor[RGB]{232,191,203} 0.65} \\
ICA & {\cellcolor[RGB]{182,212,225} 0.11} & {\cellcolor[RGB]{173,215,229} 0.02} & {\cellcolor[RGB]{212,199,212} 0.43} & {\cellcolor[RGB]{175,214,228} 0.04} & {\cellcolor[RGB]{173,215,229} 0.02} \\
Morpheus & {\cellcolor[RGB]{243,186,198} 0.77} & {\cellcolor[RGB]{222,195,207} 0.54} & {\cellcolor[RGB]{225,194,206} 0.57} & {\cellcolor[RGB]{219,196,208} 0.51} & {\cellcolor[RGB]{226,193,206} 0.58} \\
\midrule
\midrule
ASR Avg. & {\cellcolor[RGB]{219,196,208} 0.51} & {\cellcolor[RGB]{200,204,217} 0.31} & {\cellcolor[RGB]{218,197,209} 0.50} & {\cellcolor[RGB]{206,202,214} 0.37} & {\cellcolor[RGB]{208,201,214} 0.39} \\
ASR Std. & {\cellcolor[RGB]{193,207,220} 0.23} & {\cellcolor[RGB]{192,207,221} 0.22} & {\cellcolor[RGB]{179,213,227} 0.08} & {\cellcolor[RGB]{194,207,220} 0.24} & {\cellcolor[RGB]{194,207,220} 0.24} \\
\bottomrule
\end{tabular}
    }
    \label{tab:white_box}
    }
\end{table}

Our framework encompasses 19 attack methods, 29 defense methods, and 19 evaluation methods to comprehensively assess LLM safety. To compare model performance against different attack methods, as shown in Table \ref{tab:black_box} and Table \ref{tab:white_box}, we calculate the ASR for each model under various jailbreaking attack methods. White-box models typically exhibit higher ASR compared to black-box models. Through detailed analysis of the samples, we find that this phenomenon occurs because certain attack samples trigger the content moderation modules of black-box models, thereby preventing the successful generation of harmful content and consequently reducing the ASR. Among the black-box model category, OpenAI models (\textit{e.g.} OpenAI-o1 and GPT-5) and Claude-3.5 demonstrate relatively low ASR, with specific values of 0.21, 0.28, and 0.11, respectively, indicating their strong adversarial robustness (likely attributed to sophisticated content moderation modules). In contrast, other models, such as Grok-3, exhibit significantly higher ASRs. Furthermore, from the perspective of attack resistance stability (ASR standard deviation), Claude-3.5 demonstrates consistently high defensive performance when confronted with various attack methods, whereas, conversely, DeepSeek-R1 exhibits widespread security vulnerabilities.

\subsection{Attack-evaluation Results and Analysis}
\begin{table}[tbp]
    \small
    \centering
    \setlength{\tabcolsep}{4pt}
    {
    \renewcommand\arraystretch{1.1}
    \caption{ASR among evaluators regarding the outcomes of attacking Vicuna-7B (Part 1).}
    \resizebox{\linewidth}{!}{
        \begin{tabular}[c]{l|cccccccccc}
\toprule
{Evaluator} & ArtPrompt & AutoDAN & Cipher & DeepInception & Dra & GCG & GPTFUZZER & ICA & Jailbroken & LAA \\
\midrule
GPTFuzzer-CLS & {\cellcolor[RGB]{179,213,227} 0.08} & {\cellcolor[RGB]{214,198,211} 0.50} & {\cellcolor[RGB]{188,209,222} 0.19} & {\cellcolor[RGB]{221,196,208} 0.58} & {\cellcolor[RGB]{232,191,203} 0.72} & {\cellcolor[RGB]{181,212,226} 0.10} & {\cellcolor[RGB]{227,193,205} 0.66} & {\cellcolor[RGB]{181,212,226} 0.10} & {\cellcolor[RGB]{179,213,227} 0.08} & {\cellcolor[RGB]{224,194,206} 0.62} \\
Claude-Sonnet-4 & {\cellcolor[RGB]{197,205,218} 0.30} & {\cellcolor[RGB]{215,198,210} 0.51} & {\cellcolor[RGB]{190,208,222} 0.21} & {\cellcolor[RGB]{252,183,194} 0.96} & {\cellcolor[RGB]{250,183,194} 0.94} & {\cellcolor[RGB]{211,200,212} 0.46} & {\cellcolor[RGB]{245,185,197} 0.88} & {\cellcolor[RGB]{182,211,225} 0.12} & {\cellcolor[RGB]{183,211,225} 0.13} & {\cellcolor[RGB]{247,185,196} 0.90} \\
DeepSeek-R1 & {\cellcolor[RGB]{191,208,221} 0.22} & {\cellcolor[RGB]{215,198,210} 0.51} & {\cellcolor[RGB]{189,209,222} 0.20} & {\cellcolor[RGB]{230,191,203} 0.70} & {\cellcolor[RGB]{235,189,201} 0.76} & {\cellcolor[RGB]{222,195,207} 0.60} & {\cellcolor[RGB]{243,186,198} 0.85} & {\cellcolor[RGB]{181,212,226} 0.10} & {\cellcolor[RGB]{184,211,224} 0.14} & {\cellcolor[RGB]{239,188,200} 0.80} \\
DeepSeek-V3 & {\cellcolor[RGB]{192,207,221} 0.24} & {\cellcolor[RGB]{215,198,210} 0.51} & {\cellcolor[RGB]{189,209,222} 0.20} & {\cellcolor[RGB]{237,189,200} 0.78} & {\cellcolor[RGB]{249,184,195} 0.92} & {\cellcolor[RGB]{248,184,195} 0.91} & {\cellcolor[RGB]{245,185,197} 0.88} & {\cellcolor[RGB]{182,212,225} 0.11} & {\cellcolor[RGB]{183,211,225} 0.13} & {\cellcolor[RGB]{247,185,196} 0.90} \\
Doubao & {\cellcolor[RGB]{189,209,222} 0.20} & {\cellcolor[RGB]{214,198,211} 0.50} & {\cellcolor[RGB]{189,209,222} 0.20} & {\cellcolor[RGB]{236,189,201} 0.77} & {\cellcolor[RGB]{245,186,197} 0.87} & {\cellcolor[RGB]{198,205,218} 0.31} & {\cellcolor[RGB]{241,187,198} 0.83} & {\cellcolor[RGB]{182,211,225} 0.12} & {\cellcolor[RGB]{182,211,225} 0.12} & {\cellcolor[RGB]{239,188,200} 0.80} \\
GPT-5 & {\cellcolor[RGB]{192,208,221} 0.23} & {\cellcolor[RGB]{214,198,211} 0.50} & {\cellcolor[RGB]{190,208,222} 0.21} & {\cellcolor[RGB]{240,188,199} 0.81} & {\cellcolor[RGB]{245,185,197} 0.88} & {\cellcolor[RGB]{219,196,209} 0.56} & {\cellcolor[RGB]{245,186,197} 0.87} & {\cellcolor[RGB]{182,211,225} 0.12} & {\cellcolor[RGB]{182,211,225} 0.12} & {\cellcolor[RGB]{242,187,198} 0.84} \\
GPT-4 & {\cellcolor[RGB]{197,206,219} 0.29} & {\cellcolor[RGB]{214,198,211} 0.50} & {\cellcolor[RGB]{190,208,222} 0.21} & {\cellcolor[RGB]{240,188,199} 0.81} & {\cellcolor[RGB]{250,184,195} 0.93} & {\cellcolor[RGB]{245,185,197} 0.88} & {\cellcolor[RGB]{249,184,195} 0.92} & {\cellcolor[RGB]{182,211,225} 0.12} & {\cellcolor[RGB]{183,211,225} 0.13} & {\cellcolor[RGB]{247,185,196} 0.90} \\
Gemini & {\cellcolor[RGB]{192,207,221} 0.24} & {\cellcolor[RGB]{215,198,210} 0.51} & {\cellcolor[RGB]{189,209,222} 0.20} & {\cellcolor[RGB]{238,188,200} 0.79} & {\cellcolor[RGB]{247,185,196} 0.90} & {\cellcolor[RGB]{227,193,205} 0.66} & {\cellcolor[RGB]{245,185,197} 0.88} & {\cellcolor[RGB]{182,212,225} 0.11} & {\cellcolor[RGB]{182,211,225} 0.12} & {\cellcolor[RGB]{245,186,197} 0.87} \\
Grok-3 & {\cellcolor[RGB]{193,207,220} 0.25} & {\cellcolor[RGB]{214,198,211} 0.50} & {\cellcolor[RGB]{190,208,222} 0.21} & {\cellcolor[RGB]{239,188,200} 0.80} & {\cellcolor[RGB]{246,185,196} 0.89} & {\cellcolor[RGB]{222,195,207} 0.60} & {\cellcolor[RGB]{245,186,197} 0.87} & {\cellcolor[RGB]{182,212,225} 0.11} & {\cellcolor[RGB]{187,209,223} 0.18} & {\cellcolor[RGB]{244,186,197} 0.86} \\
HarmBench-CLS & {\cellcolor[RGB]{177,213,227} 0.06} & {\cellcolor[RGB]{214,198,211} 0.50} & {\cellcolor[RGB]{189,209,222} 0.20} & {\cellcolor[RGB]{238,188,200} 0.79} & {\cellcolor[RGB]{247,185,196} 0.90} & {\cellcolor[RGB]{228,192,204} 0.67} & {\cellcolor[RGB]{245,185,197} 0.88} & {\cellcolor[RGB]{182,211,225} 0.12} & {\cellcolor[RGB]{178,213,227} 0.07} & {\cellcolor[RGB]{244,186,197} 0.86} \\
Kimi & {\cellcolor[RGB]{191,208,221} 0.22} & {\cellcolor[RGB]{214,198,211} 0.50} & {\cellcolor[RGB]{189,209,222} 0.20} & {\cellcolor[RGB]{240,187,199} 0.82} & {\cellcolor[RGB]{251,183,194} 0.95} & {\cellcolor[RGB]{245,185,197} 0.88} & {\cellcolor[RGB]{243,186,198} 0.85} & {\cellcolor[RGB]{182,212,225} 0.11} & {\cellcolor[RGB]{183,211,225} 0.13} & {\cellcolor[RGB]{251,183,194} 0.95} \\
Llama-3.1-Instruct & {\cellcolor[RGB]{192,207,221} 0.24} & {\cellcolor[RGB]{214,198,211} 0.50} & {\cellcolor[RGB]{185,210,224} 0.15} & {\cellcolor[RGB]{240,188,199} 0.81} & {\cellcolor[RGB]{247,185,196} 0.90} & {\cellcolor[RGB]{231,191,203} 0.71} & {\cellcolor[RGB]{245,186,197} 0.87} & {\cellcolor[RGB]{182,212,225} 0.11} & {\cellcolor[RGB]{183,211,225} 0.13} & {\cellcolor[RGB]{245,185,197} 0.88} \\
LlamaGuard-3 & {\cellcolor[RGB]{201,204,217} 0.34} & {\cellcolor[RGB]{214,198,211} 0.50} & {\cellcolor[RGB]{188,209,222} 0.19} & {\cellcolor[RGB]{242,187,198} 0.84} & {\cellcolor[RGB]{247,185,196} 0.90} & {\cellcolor[RGB]{248,184,195} 0.91} & {\cellcolor[RGB]{240,187,199} 0.82} & {\cellcolor[RGB]{182,212,225} 0.11} & {\cellcolor[RGB]{184,211,224} 0.14} & {\cellcolor[RGB]{232,191,203} 0.72} \\
RADAR & {\cellcolor[RGB]{180,212,226} 0.09} & {\cellcolor[RGB]{214,198,211} 0.50} & {\cellcolor[RGB]{182,211,225} 0.12} & {\cellcolor[RGB]{186,210,224} 0.16} & {\cellcolor[RGB]{195,206,219} 0.27} & {\cellcolor[RGB]{182,212,225} 0.11} & {\cellcolor[RGB]{180,212,226} 0.09} & {\cellcolor[RGB]{185,210,224} 0.15} & {\cellcolor[RGB]{181,212,226} 0.10} & {\cellcolor[RGB]{206,201,214} 0.41} \\
PatternMatch & {\cellcolor[RGB]{219,196,209} 0.56} & {\cellcolor[RGB]{214,198,211} 0.50} & {\cellcolor[RGB]{184,211,224} 0.14} & {\cellcolor[RGB]{230,191,203} 0.70} & {\cellcolor[RGB]{227,193,205} 0.66} & {\cellcolor[RGB]{250,184,195} 0.93} & {\cellcolor[RGB]{199,205,218} 0.32} & {\cellcolor[RGB]{182,212,225} 0.11} & {\cellcolor[RGB]{188,209,222} 0.19} & {\cellcolor[RGB]{200,204,217} 0.33} \\
PrefixMatch & {\cellcolor[RGB]{177,214,228} 0.05} & {\cellcolor[RGB]{213,199,211} 0.49} & {\cellcolor[RGB]{173,215,229} 0.01} & {\cellcolor[RGB]{225,194,206} 0.63} & {\cellcolor[RGB]{229,192,204} 0.68} & {\cellcolor[RGB]{173,216,230} 0.00} & {\cellcolor[RGB]{216,197,210} 0.53} & {\cellcolor[RGB]{177,213,227} 0.06} & {\cellcolor[RGB]{173,215,229} 0.01} & {\cellcolor[RGB]{221,195,207} 0.59} \\
Qwen-2.5-Instruct & {\cellcolor[RGB]{187,209,223} 0.18} & {\cellcolor[RGB]{215,198,210} 0.51} & {\cellcolor[RGB]{183,211,225} 0.13} & {\cellcolor[RGB]{199,205,218} 0.32} & {\cellcolor[RGB]{233,190,202} 0.73} & {\cellcolor[RGB]{233,190,202} 0.73} & {\cellcolor[RGB]{239,188,200} 0.80} & {\cellcolor[RGB]{178,213,227} 0.07} & {\cellcolor[RGB]{182,212,225} 0.11} & {\cellcolor[RGB]{231,191,203} 0.71} \\
ShieldLM & {\cellcolor[RGB]{210,200,213} 0.45} & {\cellcolor[RGB]{214,198,211} 0.50} & {\cellcolor[RGB]{190,208,222} 0.21} & {\cellcolor[RGB]{246,185,196} 0.89} & {\cellcolor[RGB]{255,182,193} 0.99} & {\cellcolor[RGB]{251,183,194} 0.95} & {\cellcolor[RGB]{243,186,198} 0.85} & {\cellcolor[RGB]{182,212,225} 0.11} & {\cellcolor[RGB]{185,210,224} 0.15} & {\cellcolor[RGB]{230,191,203} 0.70} \\
ShieldGemma & {\cellcolor[RGB]{197,205,218} 0.30} & {\cellcolor[RGB]{214,198,211} 0.50} & {\cellcolor[RGB]{173,215,229} 0.01} & {\cellcolor[RGB]{174,215,229} 0.02} & {\cellcolor[RGB]{174,215,229} 0.02} & {\cellcolor[RGB]{174,215,229} 0.02} & {\cellcolor[RGB]{175,214,228} 0.03} & {\cellcolor[RGB]{180,212,226} 0.09} & {\cellcolor[RGB]{173,215,229} 0.01} & {\cellcolor[RGB]{173,215,229} 0.01} \\
\midrule
\midrule
ASR Avg. & {\cellcolor[RGB]{192,207,221} 0.24} & {\cellcolor[RGB]{214,198,211} 0.50} & {\cellcolor[RGB]{187,210,223} 0.17} & {\cellcolor[RGB]{229,192,204} 0.68} & {\cellcolor[RGB]{237,189,200} 0.78} & {\cellcolor[RGB]{221,196,208} 0.58} & {\cellcolor[RGB]{232,191,203} 0.72} & {\cellcolor[RGB]{182,212,225} 0.11} & {\cellcolor[RGB]{182,211,225} 0.12} & {\cellcolor[RGB]{232,191,203} 0.72} \\
ASR Std. & {\cellcolor[RGB]{182,211,225} 0.12} & {\cellcolor[RGB]{173,215,229} 0.01} & {\cellcolor[RGB]{177,213,227} 0.06} & {\cellcolor[RGB]{192,207,221} 0.24} & {\cellcolor[RGB]{192,207,221} 0.24} & {\cellcolor[RGB]{199,205,218} 0.32} & {\cellcolor[RGB]{195,206,219} 0.27} & {\cellcolor[RGB]{174,215,229} 0.02} & {\cellcolor[RGB]{177,214,228} 0.05} & {\cellcolor[RGB]{192,208,221} 0.23} \\
\bottomrule
\end{tabular}
    }
    \label{tab:eval_attack_part1}
    }
\end{table}

\begin{table}[htbp]
    \small
    \centering
    \setlength{\tabcolsep}{4pt}
    {
    \renewcommand\arraystretch{1.1}
    \caption{ASR among evaluators regarding the outcomes of attacking Vicuna-7B (Part 2).}
    \resizebox{\linewidth}{!}{
        \begin{tabular}[c]{l|ccccccccc}
\toprule
{Evaluator} & Morpheus & MultiLingual & Overload & PAIR & Past tense & RENE & Random Search & SCAV & TAP \\
\midrule
GPTFuzzer-CLS & {\cellcolor[RGB]{224,194,206} 0.62} & {\cellcolor[RGB]{251,183,194} 0.95} & {\cellcolor[RGB]{232,191,203} 0.72} & {\cellcolor[RGB]{197,206,219} 0.29} & {\cellcolor[RGB]{235,190,201} 0.75} & {\cellcolor[RGB]{216,197,210} 0.53} & {\cellcolor[RGB]{203,203,216} 0.37} & {\cellcolor[RGB]{237,189,200} 0.78} & {\cellcolor[RGB]{187,209,223} 0.18} \\
Claude-Sonnet-4 & {\cellcolor[RGB]{255,182,193} 0.99} & {\cellcolor[RGB]{196,206,219} 0.28} & {\cellcolor[RGB]{241,187,198} 0.83} & {\cellcolor[RGB]{253,182,193} 0.97} & {\cellcolor[RGB]{234,190,202} 0.74} & {\cellcolor[RGB]{222,195,207} 0.60} & {\cellcolor[RGB]{225,194,206} 0.63} & {\cellcolor[RGB]{243,186,198} 0.85} & {\cellcolor[RGB]{200,204,217} 0.33} \\
DeepSeek-R1 & {\cellcolor[RGB]{246,185,196} 0.89} & {\cellcolor[RGB]{191,208,221} 0.22} & {\cellcolor[RGB]{231,191,203} 0.71} & {\cellcolor[RGB]{253,182,193} 0.97} & {\cellcolor[RGB]{208,201,213} 0.43} & {\cellcolor[RGB]{210,200,213} 0.45} & {\cellcolor[RGB]{232,191,203} 0.72} & {\cellcolor[RGB]{239,188,200} 0.80} & {\cellcolor[RGB]{197,206,219} 0.29} \\
DeepSeek-V3 & {\cellcolor[RGB]{250,184,195} 0.93} & {\cellcolor[RGB]{193,207,220} 0.25} & {\cellcolor[RGB]{234,190,202} 0.74} & {\cellcolor[RGB]{254,182,193} 0.98} & {\cellcolor[RGB]{229,192,204} 0.68} & {\cellcolor[RGB]{219,196,209} 0.56} & {\cellcolor[RGB]{252,183,194} 0.96} & {\cellcolor[RGB]{240,187,199} 0.82} & {\cellcolor[RGB]{193,207,220} 0.25} \\
Doubao & {\cellcolor[RGB]{252,183,194} 0.96} & {\cellcolor[RGB]{194,207,220} 0.26} & {\cellcolor[RGB]{237,189,200} 0.78} & {\cellcolor[RGB]{253,182,193} 0.97} & {\cellcolor[RGB]{230,192,204} 0.69} & {\cellcolor[RGB]{214,198,211} 0.50} & {\cellcolor[RGB]{216,198,210} 0.52} & {\cellcolor[RGB]{240,187,199} 0.82} & {\cellcolor[RGB]{195,206,219} 0.27} \\
GPT-5 & {\cellcolor[RGB]{250,183,194} 0.94} & {\cellcolor[RGB]{173,216,230} 0.00} & {\cellcolor[RGB]{235,189,201} 0.76} & {\cellcolor[RGB]{253,182,193} 0.97} & {\cellcolor[RGB]{225,194,206} 0.63} & {\cellcolor[RGB]{216,197,210} 0.53} & {\cellcolor[RGB]{231,191,203} 0.71} & {\cellcolor[RGB]{240,187,199} 0.82} & {\cellcolor[RGB]{196,206,219} 0.28} \\
GPT-4 & {\cellcolor[RGB]{253,182,193} 0.97} & {\cellcolor[RGB]{196,206,219} 0.28} & {\cellcolor[RGB]{238,188,200} 0.79} & {\cellcolor[RGB]{252,183,194} 0.96} & {\cellcolor[RGB]{225,194,206} 0.63} & {\cellcolor[RGB]{216,197,210} 0.53} & {\cellcolor[RGB]{249,184,195} 0.92} & {\cellcolor[RGB]{243,186,198} 0.85} & {\cellcolor[RGB]{196,206,219} 0.28} \\
Gemini & {\cellcolor[RGB]{252,183,194} 0.96} & {\cellcolor[RGB]{189,209,222} 0.20} & {\cellcolor[RGB]{235,189,201} 0.76} & {\cellcolor[RGB]{252,183,194} 0.96} & {\cellcolor[RGB]{227,193,205} 0.66} & {\cellcolor[RGB]{216,197,210} 0.53} & {\cellcolor[RGB]{236,189,201} 0.77} & {\cellcolor[RGB]{241,187,198} 0.83} & {\cellcolor[RGB]{194,207,220} 0.26} \\
Grok-3 & {\cellcolor[RGB]{252,183,194} 0.96} & {\cellcolor[RGB]{197,205,218} 0.30} & {\cellcolor[RGB]{237,189,200} 0.78} & {\cellcolor[RGB]{254,182,193} 0.98} & {\cellcolor[RGB]{227,193,205} 0.66} & {\cellcolor[RGB]{216,198,210} 0.52} & {\cellcolor[RGB]{234,190,202} 0.74} & {\cellcolor[RGB]{240,187,199} 0.82} & {\cellcolor[RGB]{199,205,218} 0.32} \\
HarmBench-CLS & {\cellcolor[RGB]{251,183,194} 0.95} & {\cellcolor[RGB]{186,210,224} 0.16} & {\cellcolor[RGB]{228,192,204} 0.67} & {\cellcolor[RGB]{187,209,223} 0.18} & {\cellcolor[RGB]{198,205,218} 0.31} & {\cellcolor[RGB]{201,204,217} 0.34} & {\cellcolor[RGB]{236,189,201} 0.77} & {\cellcolor[RGB]{235,189,201} 0.76} & {\cellcolor[RGB]{185,210,224} 0.15} \\
Kimi & {\cellcolor[RGB]{251,183,194} 0.95} & {\cellcolor[RGB]{193,207,220} 0.25} & {\cellcolor[RGB]{237,189,200} 0.78} & {\cellcolor[RGB]{252,183,194} 0.96} & {\cellcolor[RGB]{226,193,205} 0.65} & {\cellcolor[RGB]{217,197,209} 0.54} & {\cellcolor[RGB]{247,185,196} 0.90} & {\cellcolor[RGB]{240,187,199} 0.82} & {\cellcolor[RGB]{193,207,220} 0.25} \\
Llama-3.1-Instruct & {\cellcolor[RGB]{252,183,194} 0.96} & {\cellcolor[RGB]{185,210,224} 0.15} & {\cellcolor[RGB]{210,200,213} 0.45} & {\cellcolor[RGB]{220,196,208} 0.57} & {\cellcolor[RGB]{214,198,211} 0.50} & {\cellcolor[RGB]{199,205,218} 0.32} & {\cellcolor[RGB]{254,182,193} 0.98} & {\cellcolor[RGB]{211,200,212} 0.46} & {\cellcolor[RGB]{191,208,221} 0.22} \\
LlamaGuard-3 & {\cellcolor[RGB]{249,184,195} 0.92} & {\cellcolor[RGB]{189,209,222} 0.20} & {\cellcolor[RGB]{238,188,200} 0.79} & {\cellcolor[RGB]{250,183,194} 0.94} & {\cellcolor[RGB]{229,192,204} 0.68} & {\cellcolor[RGB]{220,196,208} 0.57} & {\cellcolor[RGB]{253,182,193} 0.97} & {\cellcolor[RGB]{239,188,200} 0.80} & {\cellcolor[RGB]{194,207,220} 0.26} \\
RADAR & {\cellcolor[RGB]{187,209,223} 0.18} & {\cellcolor[RGB]{204,202,215} 0.38} & {\cellcolor[RGB]{185,210,224} 0.15} & {\cellcolor[RGB]{174,215,229} 0.02} & {\cellcolor[RGB]{186,210,224} 0.16} & {\cellcolor[RGB]{193,207,220} 0.25} & {\cellcolor[RGB]{187,209,223} 0.18} & {\cellcolor[RGB]{186,210,224} 0.16} & {\cellcolor[RGB]{182,212,225} 0.11} \\
PatternMatch & {\cellcolor[RGB]{243,186,198} 0.85} & {\cellcolor[RGB]{193,207,220} 0.25} & {\cellcolor[RGB]{224,194,206} 0.62} & {\cellcolor[RGB]{242,187,198} 0.84} & {\cellcolor[RGB]{226,193,205} 0.65} & {\cellcolor[RGB]{225,194,206} 0.63} & {\cellcolor[RGB]{235,189,201} 0.76} & {\cellcolor[RGB]{230,191,203} 0.70} & {\cellcolor[RGB]{205,202,215} 0.39} \\
PrefixMatch & {\cellcolor[RGB]{175,214,228} 0.03} & {\cellcolor[RGB]{192,207,221} 0.24} & {\cellcolor[RGB]{214,198,211} 0.50} & {\cellcolor[RGB]{221,195,207} 0.59} & {\cellcolor[RGB]{214,198,211} 0.50} & {\cellcolor[RGB]{210,200,213} 0.45} & {\cellcolor[RGB]{174,215,229} 0.02} & {\cellcolor[RGB]{216,197,210} 0.53} & {\cellcolor[RGB]{195,206,219} 0.27} \\
Qwen-2.5-Instruct & {\cellcolor[RGB]{221,195,207} 0.59} & {\cellcolor[RGB]{182,211,225} 0.12} & {\cellcolor[RGB]{200,204,217} 0.33} & {\cellcolor[RGB]{252,183,194} 0.96} & {\cellcolor[RGB]{187,210,223} 0.17} & {\cellcolor[RGB]{203,203,216} 0.37} & {\cellcolor[RGB]{251,183,194} 0.95} & {\cellcolor[RGB]{221,196,208} 0.58} & {\cellcolor[RGB]{188,209,222} 0.19} \\
ShieldLM & {\cellcolor[RGB]{253,182,193} 0.97} & {\cellcolor[RGB]{195,206,219} 0.27} & {\cellcolor[RGB]{240,188,199} 0.81} & {\cellcolor[RGB]{252,183,194} 0.96} & {\cellcolor[RGB]{241,187,198} 0.83} & {\cellcolor[RGB]{225,194,206} 0.63} & {\cellcolor[RGB]{247,185,196} 0.90} & {\cellcolor[RGB]{240,187,199} 0.82} & {\cellcolor[RGB]{213,199,211} 0.49} \\
ShieldGemma & {\cellcolor[RGB]{175,214,228} 0.03} & {\cellcolor[RGB]{173,215,229} 0.01} & {\cellcolor[RGB]{175,214,228} 0.03} & {\cellcolor[RGB]{174,215,229} 0.02} & {\cellcolor[RGB]{174,215,229} 0.02} & {\cellcolor[RGB]{173,215,229} 0.01} & {\cellcolor[RGB]{174,215,229} 0.02} & {\cellcolor[RGB]{176,214,228} 0.04} & {\cellcolor[RGB]{173,216,230} 0.00} \\
\midrule
\midrule
ASR Avg. & {\cellcolor[RGB]{236,189,201} 0.77} & {\cellcolor[RGB]{193,207,220} 0.25} & {\cellcolor[RGB]{225,194,206} 0.63} & {\cellcolor[RGB]{234,190,202} 0.74} & {\cellcolor[RGB]{217,197,209} 0.54} & {\cellcolor[RGB]{211,199,212} 0.47} & {\cellcolor[RGB]{228,192,204} 0.67} & {\cellcolor[RGB]{230,192,204} 0.69} & {\cellcolor[RGB]{193,207,220} 0.25} \\
ASR Std. & {\cellcolor[RGB]{199,205,218} 0.32} & {\cellcolor[RGB]{188,209,222} 0.19} & {\cellcolor[RGB]{192,208,221} 0.23} & {\cellcolor[RGB]{201,204,217} 0.34} & {\cellcolor[RGB]{191,208,221} 0.22} & {\cellcolor[RGB]{185,210,224} 0.15} & {\cellcolor[RGB]{197,205,218} 0.30} & {\cellcolor[RGB]{192,208,221} 0.23} & {\cellcolor[RGB]{181,212,226} 0.10} \\
\bottomrule
\end{tabular}
    }
    \label{tab:eval_attack_part2}
    }
\end{table}
To investigate the impact of different evaluators on ASR measurement, we conduct systematic testing using the Vicuna-7B model under no-defense conditions. As shown in Table~\ref{tab:eval_attack_part1} and Table~\ref{tab:eval_attack_part2}, the ASRs computed by different evaluators exhibit significant disparities. Notably, the evaluation of PAIR-based attack results demonstrates the greatest inconsistency, with a standard deviation of 0.34 among evaluators. This finding reflects that the current field of large language model security evaluation still lacks unified standards in specific scenarios, with substantial disagreements among different evaluators regarding the determination of ``attack success.'' These results underscore the critical importance of establishing a standardized evaluation framework.

\subsection{Attack-defense Results and Analysis}
\begin{table}[tbp]
    \centering
    \setlength{\tabcolsep}{2pt}
    {
    \renewcommand\arraystretch{1.1}
    \caption{ASR between different defense methods and multiple attack methods on Vicuna-7b (Part1).}
    \resizebox{\linewidth}{!}{
\begin{tabular}[c]{l|cccccccccc}
\toprule
{Defense Method} & Morpheus & Overload & ICA & Random Search & LAA & DeepInception & DRA & RENE & Cipher & Jailbroken \\
\midrule
PPL & {\cellcolor[RGB]{235,190,201} 0.55} & {\cellcolor[RGB]{209,200,213} 0.33} & {\cellcolor[RGB]{173,216,230} 0.01} & {\cellcolor[RGB]{200,204,217} 0.25} & {\cellcolor[RGB]{244,186,197} 0.63} & {\cellcolor[RGB]{214,198,211} 0.37} & {\cellcolor[RGB]{194,206,220} 0.20} & {\cellcolor[RGB]{211,200,212} 0.34} & {\cellcolor[RGB]{183,211,225} 0.10} & {\cellcolor[RGB]{184,211,224} 0.11} \\
Prompt Guard & {\cellcolor[RGB]{229,192,204} 0.50} & {\cellcolor[RGB]{197,205,219} 0.22} & {\cellcolor[RGB]{174,215,229} 0.02} & {\cellcolor[RGB]{238,188,200} 0.58} & {\cellcolor[RGB]{251,183,194} 0.69} & {\cellcolor[RGB]{214,198,211} 0.37} & {\cellcolor[RGB]{199,204,218} 0.24} & {\cellcolor[RGB]{175,215,228} 0.03} & {\cellcolor[RGB]{173,216,230} 0.01} & {\cellcolor[RGB]{174,215,229} 0.02} \\
Paraphrasing & {\cellcolor[RGB]{248,184,196} 0.66} & {\cellcolor[RGB]{204,203,215} 0.28} & {\cellcolor[RGB]{175,215,228} 0.03} & {\cellcolor[RGB]{223,194,207} 0.45} & {\cellcolor[RGB]{216,197,210} 0.39} & {\cellcolor[RGB]{218,197,209} 0.40} & {\cellcolor[RGB]{186,210,223} 0.13} & {\cellcolor[RGB]{190,208,222} 0.16} & {\cellcolor[RGB]{176,214,228} 0.04} & {\cellcolor[RGB]{176,214,228} 0.04} \\
EDDF & {\cellcolor[RGB]{213,199,211} 0.36} & {\cellcolor[RGB]{196,206,219} 0.21} & {\cellcolor[RGB]{184,211,224} 0.11} & {\cellcolor[RGB]{239,188,199} 0.59} & {\cellcolor[RGB]{229,192,204} 0.50} & {\cellcolor[RGB]{201,204,216} 0.26} & {\cellcolor[RGB]{198,205,218} 0.23} & {\cellcolor[RGB]{181,212,226} 0.08} & {\cellcolor[RGB]{183,211,225} 0.10} & {\cellcolor[RGB]{176,214,228} 0.04} \\
IBProtector & {\cellcolor[RGB]{226,193,206} 0.47} & {\cellcolor[RGB]{216,197,210} 0.39} & {\cellcolor[RGB]{178,213,227} 0.06} & {\cellcolor[RGB]{207,201,214} 0.31} & {\cellcolor[RGB]{209,200,213} 0.33} & {\cellcolor[RGB]{205,202,215} 0.29} & {\cellcolor[RGB]{198,205,218} 0.23} & {\cellcolor[RGB]{216,197,210} 0.39} & {\cellcolor[RGB]{177,214,227} 0.05} & {\cellcolor[RGB]{179,213,226} 0.07} \\
Self-Defense & {\cellcolor[RGB]{234,190,202} 0.54} & {\cellcolor[RGB]{209,200,213} 0.33} & {\cellcolor[RGB]{173,216,230} 0.01} & {\cellcolor[RGB]{206,202,214} 0.30} & {\cellcolor[RGB]{251,183,194} 0.69} & {\cellcolor[RGB]{214,198,211} 0.37} & {\cellcolor[RGB]{193,207,220} 0.19} & {\cellcolor[RGB]{207,201,214} 0.31} & {\cellcolor[RGB]{184,211,224} 0.11} & {\cellcolor[RGB]{181,212,226} 0.08} \\
Erase and Check & {\cellcolor[RGB]{220,196,208} 0.42} & {\cellcolor[RGB]{193,207,220} 0.19} & {\cellcolor[RGB]{173,216,230} 0.01} & {\cellcolor[RGB]{244,186,197} 0.63} & {\cellcolor[RGB]{222,195,207} 0.44} & {\cellcolor[RGB]{189,209,222} 0.15} & {\cellcolor[RGB]{182,212,225} 0.09} & {\cellcolor[RGB]{199,204,218} 0.24} & {\cellcolor[RGB]{190,208,222} 0.16} & {\cellcolor[RGB]{175,215,228} 0.03} \\
RA-LLM & {\cellcolor[RGB]{177,214,227} 0.05} & {\cellcolor[RGB]{188,209,223} 0.14} & {\cellcolor[RGB]{174,215,229} 0.02} & {\cellcolor[RGB]{197,205,219} 0.22} & {\cellcolor[RGB]{231,191,203} 0.52} & {\cellcolor[RGB]{184,211,224} 0.11} & {\cellcolor[RGB]{178,213,227} 0.06} & {\cellcolor[RGB]{186,210,223} 0.13} & {\cellcolor[RGB]{191,208,221} 0.17} & {\cellcolor[RGB]{178,213,227} 0.06} \\
Aligner & {\cellcolor[RGB]{235,190,201} 0.55} & {\cellcolor[RGB]{199,204,218} 0.24} & {\cellcolor[RGB]{182,212,225} 0.09} & {\cellcolor[RGB]{191,208,221} 0.17} & {\cellcolor[RGB]{199,204,218} 0.24} & {\cellcolor[RGB]{201,204,216} 0.26} & {\cellcolor[RGB]{201,204,216} 0.26} & {\cellcolor[RGB]{179,213,226} 0.07} & {\cellcolor[RGB]{182,212,225} 0.09} & {\cellcolor[RGB]{179,213,226} 0.07} \\
BackTranslation & {\cellcolor[RGB]{218,197,209} 0.40} & {\cellcolor[RGB]{192,207,221} 0.18} & {\cellcolor[RGB]{178,213,227} 0.06} & {\cellcolor[RGB]{198,205,218} 0.23} & {\cellcolor[RGB]{200,204,217} 0.25} & {\cellcolor[RGB]{189,209,222} 0.15} & {\cellcolor[RGB]{189,209,222} 0.15} & {\cellcolor[RGB]{184,211,224} 0.11} & {\cellcolor[RGB]{175,215,228} 0.03} & {\cellcolor[RGB]{181,212,226} 0.08} \\
Gradient Cuff & {\cellcolor[RGB]{220,196,208} 0.42} & {\cellcolor[RGB]{209,200,213} 0.33} & {\cellcolor[RGB]{173,216,230} 0.01} & {\cellcolor[RGB]{234,190,202} 0.54} & {\cellcolor[RGB]{228,193,204} 0.49} & {\cellcolor[RGB]{224,194,206} 0.46} & {\cellcolor[RGB]{212,199,212} 0.35} & {\cellcolor[RGB]{194,206,220} 0.20} & {\cellcolor[RGB]{191,208,221} 0.17} & {\cellcolor[RGB]{182,212,225} 0.09} \\
GuardReasoner & {\cellcolor[RGB]{208,201,213} 0.32} & {\cellcolor[RGB]{186,210,223} 0.13} & {\cellcolor[RGB]{173,216,230} 0.01} & {\cellcolor[RGB]{236,189,201} 0.56} & {\cellcolor[RGB]{209,200,213} 0.33} & {\cellcolor[RGB]{177,214,227} 0.05} & {\cellcolor[RGB]{184,211,224} 0.11} & {\cellcolor[RGB]{179,213,226} 0.07} & {\cellcolor[RGB]{181,212,226} 0.08} & {\cellcolor[RGB]{173,216,230} 0.01} \\
SelfReminder & {\cellcolor[RGB]{239,188,199} 0.59} & {\cellcolor[RGB]{213,199,211} 0.36} & {\cellcolor[RGB]{181,212,226} 0.08} & {\cellcolor[RGB]{242,187,198} 0.61} & {\cellcolor[RGB]{244,186,197} 0.63} & {\cellcolor[RGB]{196,206,219} 0.21} & {\cellcolor[RGB]{173,216,230} 0.01} & {\cellcolor[RGB]{181,212,226} 0.08} & {\cellcolor[RGB]{176,214,228} 0.04} & {\cellcolor[RGB]{175,215,228} 0.03} \\
ICD & {\cellcolor[RGB]{177,214,227} 0.05} & {\cellcolor[RGB]{188,209,223} 0.14} & {\cellcolor[RGB]{173,216,230} 0.01} & {\cellcolor[RGB]{199,204,218} 0.24} & {\cellcolor[RGB]{198,205,218} 0.23} & {\cellcolor[RGB]{182,212,225} 0.09} & {\cellcolor[RGB]{173,216,230} 0.01} & {\cellcolor[RGB]{177,214,227} 0.05} & {\cellcolor[RGB]{176,214,228} 0.04} & {\cellcolor[RGB]{175,215,228} 0.03} \\
DRO & {\cellcolor[RGB]{230,192,203} 0.51} & {\cellcolor[RGB]{204,203,215} 0.28} & {\cellcolor[RGB]{183,211,225} 0.10} & {\cellcolor[RGB]{246,185,196} 0.65} & {\cellcolor[RGB]{241,187,199} 0.60} & {\cellcolor[RGB]{226,193,206} 0.47} & {\cellcolor[RGB]{229,192,204} 0.50} & {\cellcolor[RGB]{186,210,223} 0.13} & {\cellcolor[RGB]{177,214,227} 0.05} & {\cellcolor[RGB]{177,214,227} 0.05} \\
GoalPriority & {\cellcolor[RGB]{184,211,224} 0.11} & {\cellcolor[RGB]{204,203,215} 0.28} & {\cellcolor[RGB]{176,214,228} 0.04} & {\cellcolor[RGB]{197,205,219} 0.22} & {\cellcolor[RGB]{239,188,199} 0.59} & {\cellcolor[RGB]{207,201,214} 0.31} & {\cellcolor[RGB]{215,198,210} 0.38} & {\cellcolor[RGB]{192,207,221} 0.18} & {\cellcolor[RGB]{174,215,229} 0.02} & {\cellcolor[RGB]{173,216,230} 0.01} \\
SmoothLLM & {\cellcolor[RGB]{226,193,206} 0.47} & {\cellcolor[RGB]{204,203,215} 0.28} & {\cellcolor[RGB]{173,216,230} 0.01} & {\cellcolor[RGB]{201,204,216} 0.26} & {\cellcolor[RGB]{194,206,220} 0.20} & {\cellcolor[RGB]{183,211,225} 0.10} & {\cellcolor[RGB]{201,204,216} 0.26} & {\cellcolor[RGB]{212,199,212} 0.35} & {\cellcolor[RGB]{181,212,226} 0.08} & {\cellcolor[RGB]{184,211,224} 0.11} \\
SafeDecoding & {\cellcolor[RGB]{228,193,204} 0.49} & {\cellcolor[RGB]{205,202,215} 0.29} & {\cellcolor[RGB]{184,211,224} 0.11} & {\cellcolor[RGB]{238,188,200} 0.58} & {\cellcolor[RGB]{245,185,197} 0.64} & {\cellcolor[RGB]{223,194,207} 0.45} & {\cellcolor[RGB]{218,197,209} 0.40} & {\cellcolor[RGB]{194,206,220} 0.20} & {\cellcolor[RGB]{188,209,223} 0.14} & {\cellcolor[RGB]{181,212,226} 0.08} \\
RPO & {\cellcolor[RGB]{216,197,210} 0.39} & {\cellcolor[RGB]{206,202,214} 0.30} & {\cellcolor[RGB]{173,216,230} 0.01} & {\cellcolor[RGB]{237,189,200} 0.57} & {\cellcolor[RGB]{242,187,198} 0.61} & {\cellcolor[RGB]{199,204,218} 0.24} & {\cellcolor[RGB]{216,197,210} 0.39} & {\cellcolor[RGB]{183,211,225} 0.10} & {\cellcolor[RGB]{178,213,227} 0.06} & {\cellcolor[RGB]{181,212,226} 0.08} \\
RePE & {\cellcolor[RGB]{226,193,206} 0.47} & {\cellcolor[RGB]{233,191,202} 0.53} & {\cellcolor[RGB]{173,216,230} 0.01} & {\cellcolor[RGB]{241,187,199} 0.60} & {\cellcolor[RGB]{244,186,197} 0.63} & {\cellcolor[RGB]{221,195,208} 0.43} & {\cellcolor[RGB]{224,194,206} 0.46} & {\cellcolor[RGB]{220,196,208} 0.42} & {\cellcolor[RGB]{177,214,227} 0.05} & {\cellcolor[RGB]{173,216,230} 0.01} \\
GradSafe & {\cellcolor[RGB]{228,193,204} 0.49} & {\cellcolor[RGB]{208,201,213} 0.32} & {\cellcolor[RGB]{173,216,230} 0.01} & {\cellcolor[RGB]{237,189,200} 0.57} & {\cellcolor[RGB]{244,186,197} 0.63} & {\cellcolor[RGB]{213,199,211} 0.36} & {\cellcolor[RGB]{174,215,229} 0.02} & {\cellcolor[RGB]{179,213,226} 0.07} & {\cellcolor[RGB]{186,210,223} 0.13} & {\cellcolor[RGB]{175,215,228} 0.03} \\
AVGAN & {\cellcolor[RGB]{191,208,221} 0.17} & {\cellcolor[RGB]{177,214,227} 0.05} & {\cellcolor[RGB]{184,211,224} 0.11} & {\cellcolor[RGB]{208,201,213} 0.32} & {\cellcolor[RGB]{219,196,209} 0.41} & {\cellcolor[RGB]{211,200,212} 0.34} & {\cellcolor[RGB]{204,203,215} 0.28} & {\cellcolor[RGB]{177,214,227} 0.05} & {\cellcolor[RGB]{190,208,222} 0.16} & {\cellcolor[RGB]{175,215,228} 0.03} \\
JBShield & {\cellcolor[RGB]{215,198,210} 0.38} & {\cellcolor[RGB]{208,201,213} 0.32} & {\cellcolor[RGB]{176,214,228} 0.04} & {\cellcolor[RGB]{242,187,198} 0.61} & {\cellcolor[RGB]{255,182,193} 0.72} & {\cellcolor[RGB]{228,193,204} 0.49} & {\cellcolor[RGB]{218,197,209} 0.40} & {\cellcolor[RGB]{193,207,220} 0.19} & {\cellcolor[RGB]{185,210,224} 0.12} & {\cellcolor[RGB]{177,214,227} 0.05} \\
RAIN & {\cellcolor[RGB]{221,195,208} 0.43} & {\cellcolor[RGB]{218,197,209} 0.40} & {\cellcolor[RGB]{176,214,228} 0.04} & {\cellcolor[RGB]{234,190,202} 0.54} & {\cellcolor[RGB]{250,183,195} 0.68} & {\cellcolor[RGB]{230,192,203} 0.51} & {\cellcolor[RGB]{230,192,203} 0.51} & {\cellcolor[RGB]{218,197,209} 0.40} & {\cellcolor[RGB]{179,213,226} 0.07} & {\cellcolor[RGB]{173,216,230} 0.01} \\
Backdoor Alignment & {\cellcolor[RGB]{219,196,209} 0.41} & {\cellcolor[RGB]{186,210,223} 0.13} & {\cellcolor[RGB]{182,212,225} 0.09} & {\cellcolor[RGB]{201,204,216} 0.26} & {\cellcolor[RGB]{193,207,220} 0.19} & {\cellcolor[RGB]{175,215,228} 0.03} & {\cellcolor[RGB]{176,214,228} 0.04} & {\cellcolor[RGB]{183,211,225} 0.10} & {\cellcolor[RGB]{177,214,227} 0.05} & {\cellcolor[RGB]{181,212,226} 0.08} \\
DELMAN & {\cellcolor[RGB]{220,196,208} 0.42} & {\cellcolor[RGB]{214,198,211} 0.37} & {\cellcolor[RGB]{173,216,230} 0.01} & {\cellcolor[RGB]{226,193,206} 0.47} & {\cellcolor[RGB]{218,197,209} 0.40} & {\cellcolor[RGB]{224,194,206} 0.46} & {\cellcolor[RGB]{221,195,208} 0.43} & {\cellcolor[RGB]{215,198,210} 0.38} & {\cellcolor[RGB]{179,213,226} 0.07} & {\cellcolor[RGB]{175,215,228} 0.03} \\
Safety-Tuned & {\cellcolor[RGB]{221,195,208} 0.43} & {\cellcolor[RGB]{209,200,213} 0.33} & {\cellcolor[RGB]{177,214,227} 0.05} & {\cellcolor[RGB]{213,199,211} 0.36} & {\cellcolor[RGB]{242,187,198} 0.61} & {\cellcolor[RGB]{201,204,216} 0.26} & {\cellcolor[RGB]{190,208,222} 0.16} & {\cellcolor[RGB]{207,201,214} 0.31} & {\cellcolor[RGB]{181,212,226} 0.08} & {\cellcolor[RGB]{173,216,230} 0.01} \\
Layer-AdvPatcher & {\cellcolor[RGB]{248,184,196} 0.66} & {\cellcolor[RGB]{220,196,208} 0.42} & {\cellcolor[RGB]{173,216,230} 0.01} & {\cellcolor[RGB]{249,184,195} 0.67} & {\cellcolor[RGB]{251,183,194} 0.69} & {\cellcolor[RGB]{250,183,195} 0.68} & {\cellcolor[RGB]{231,191,203} 0.52} & {\cellcolor[RGB]{222,195,207} 0.44} & {\cellcolor[RGB]{179,213,226} 0.07} & {\cellcolor[RGB]{174,215,229} 0.02} \\
C-advipo & {\cellcolor[RGB]{173,216,230} 0.01} & {\cellcolor[RGB]{206,202,214} 0.30} & {\cellcolor[RGB]{179,213,226} 0.07} & {\cellcolor[RGB]{218,197,209} 0.40} & {\cellcolor[RGB]{213,199,211} 0.36} & {\cellcolor[RGB]{205,202,215} 0.29} & {\cellcolor[RGB]{235,190,201} 0.55} & {\cellcolor[RGB]{188,209,223} 0.14} & {\cellcolor[RGB]{173,216,230} 0.01} & {\cellcolor[RGB]{182,212,225} 0.09} \\
\bottomrule
\end{tabular}
    }
    \label{tab:attack_defense_a}
    }
\end{table}

\begin{table}[tbp]
    \centering
    \setlength{\tabcolsep}{2pt}
    {
    \renewcommand\arraystretch{1.1}
    \caption{ASR between different defense methods and multiple attack methods on Vicuna-7b (Part2).}
    \resizebox{\linewidth}{!}{
\begin{tabular}[c]{l|ccccccccc}
\toprule
{Defense Method} & MultiLingual & GPTFUZZER & AutoDAN & PAIR & SCAV & TAP & GCG & ArtPrompt & Past tense \\
\midrule
PPL & {\cellcolor[RGB]{187,209,223} 0.12} & {\cellcolor[RGB]{190,208,222} 0.14} & {\cellcolor[RGB]{216,197,210} 0.34} & {\cellcolor[RGB]{241,187,198} 0.53} & {\cellcolor[RGB]{235,190,201} 0.48} & {\cellcolor[RGB]{199,205,218} 0.21} & {\cellcolor[RGB]{175,214,228} 0.03} & {\cellcolor[RGB]{187,209,223} 0.12} & {\cellcolor[RGB]{219,196,209} 0.36} \\
Prompt Guard & {\cellcolor[RGB]{173,216,230} 0.01} & {\cellcolor[RGB]{202,203,216} 0.23} & {\cellcolor[RGB]{211,200,212} 0.30} & {\cellcolor[RGB]{173,216,230} 0.01} & {\cellcolor[RGB]{215,198,210} 0.33} & {\cellcolor[RGB]{186,210,224} 0.11} & {\cellcolor[RGB]{174,215,229} 0.02} & {\cellcolor[RGB]{174,215,229} 0.02} & {\cellcolor[RGB]{224,194,206} 0.40} \\
Paraphrasing & {\cellcolor[RGB]{194,207,220} 0.17} & {\cellcolor[RGB]{207,201,214} 0.27} & {\cellcolor[RGB]{208,201,213} 0.28} & {\cellcolor[RGB]{223,195,207} 0.39} & {\cellcolor[RGB]{221,195,207} 0.38} & {\cellcolor[RGB]{194,207,220} 0.17} & {\cellcolor[RGB]{214,199,211} 0.32} & {\cellcolor[RGB]{202,203,216} 0.23} & {\cellcolor[RGB]{228,192,204} 0.43} \\
EDDF & {\cellcolor[RGB]{187,209,223} 0.12} & {\cellcolor[RGB]{190,208,222} 0.14} & {\cellcolor[RGB]{211,200,212} 0.30} & {\cellcolor[RGB]{214,199,211} 0.32} & {\cellcolor[RGB]{208,201,213} 0.28} & {\cellcolor[RGB]{200,204,217} 0.22} & {\cellcolor[RGB]{208,201,213} 0.28} & {\cellcolor[RGB]{202,203,216} 0.23} & {\cellcolor[RGB]{233,190,202} 0.47} \\
IBProtector & {\cellcolor[RGB]{200,204,217} 0.22} & {\cellcolor[RGB]{207,201,214} 0.27} & {\cellcolor[RGB]{194,207,220} 0.17} & {\cellcolor[RGB]{190,208,222} 0.14} & {\cellcolor[RGB]{233,190,202} 0.47} & {\cellcolor[RGB]{182,212,225} 0.08} & {\cellcolor[RGB]{227,193,205} 0.42} & {\cellcolor[RGB]{182,212,225} 0.08} & {\cellcolor[RGB]{231,191,203} 0.45} \\
Self-Defense & {\cellcolor[RGB]{199,205,218} 0.21} & {\cellcolor[RGB]{180,212,226} 0.07} & {\cellcolor[RGB]{208,201,213} 0.28} & {\cellcolor[RGB]{208,201,213} 0.28} & {\cellcolor[RGB]{195,206,219} 0.18} & {\cellcolor[RGB]{187,209,223} 0.12} & {\cellcolor[RGB]{188,209,222} 0.13} & {\cellcolor[RGB]{191,208,221} 0.15} & {\cellcolor[RGB]{233,190,202} 0.47} \\
Erase and Check & {\cellcolor[RGB]{188,209,222} 0.13} & {\cellcolor[RGB]{195,206,219} 0.18} & {\cellcolor[RGB]{206,202,215} 0.26} & {\cellcolor[RGB]{203,203,216} 0.24} & {\cellcolor[RGB]{186,210,224} 0.11} & {\cellcolor[RGB]{184,211,224} 0.10} & {\cellcolor[RGB]{174,215,229} 0.02} & {\cellcolor[RGB]{179,213,227} 0.06} & {\cellcolor[RGB]{233,190,202} 0.47} \\
RA-LLM & {\cellcolor[RGB]{195,206,219} 0.18} & {\cellcolor[RGB]{182,212,225} 0.08} & {\cellcolor[RGB]{191,208,221} 0.15} & {\cellcolor[RGB]{175,214,228} 0.03} & {\cellcolor[RGB]{173,216,230} 0.01} & {\cellcolor[RGB]{183,211,225} 0.09} & {\cellcolor[RGB]{174,215,229} 0.02} & {\cellcolor[RGB]{175,214,228} 0.03} & {\cellcolor[RGB]{217,197,209} 0.35} \\
Aligner & {\cellcolor[RGB]{183,211,225} 0.09} & {\cellcolor[RGB]{176,214,228} 0.04} & {\cellcolor[RGB]{190,208,222} 0.14} & {\cellcolor[RGB]{190,208,222} 0.14} & {\cellcolor[RGB]{199,205,218} 0.21} & {\cellcolor[RGB]{186,210,224} 0.11} & {\cellcolor[RGB]{179,213,227} 0.06} & {\cellcolor[RGB]{187,209,223} 0.12} & {\cellcolor[RGB]{229,192,204} 0.44} \\
BackTranslation & {\cellcolor[RGB]{194,207,220} 0.17} & {\cellcolor[RGB]{183,211,225} 0.09} & {\cellcolor[RGB]{207,201,214} 0.27} & {\cellcolor[RGB]{206,202,215} 0.26} & {\cellcolor[RGB]{187,209,223} 0.12} & {\cellcolor[RGB]{192,207,221} 0.16} & {\cellcolor[RGB]{184,211,224} 0.10} & {\cellcolor[RGB]{190,208,222} 0.14} & {\cellcolor[RGB]{237,189,200} 0.50} \\
Gradient Cuff & {\cellcolor[RGB]{204,202,215} 0.25} & {\cellcolor[RGB]{203,203,216} 0.24} & {\cellcolor[RGB]{198,205,218} 0.20} & {\cellcolor[RGB]{196,206,219} 0.19} & {\cellcolor[RGB]{195,206,219} 0.18} & {\cellcolor[RGB]{203,203,216} 0.24} & {\cellcolor[RGB]{179,213,227} 0.06} & {\cellcolor[RGB]{200,204,217} 0.22} & {\cellcolor[RGB]{228,192,204} 0.43} \\
GuardReasoner & {\cellcolor[RGB]{180,212,226} 0.07} & {\cellcolor[RGB]{187,209,223} 0.12} & {\cellcolor[RGB]{199,205,218} 0.21} & {\cellcolor[RGB]{186,210,224} 0.11} & {\cellcolor[RGB]{195,206,219} 0.18} & {\cellcolor[RGB]{195,206,219} 0.18} & {\cellcolor[RGB]{175,214,228} 0.03} & {\cellcolor[RGB]{184,211,224} 0.10} & {\cellcolor[RGB]{232,191,203} 0.46} \\
SelfReminder & {\cellcolor[RGB]{184,211,224} 0.10} & {\cellcolor[RGB]{196,206,219} 0.19} & {\cellcolor[RGB]{202,203,216} 0.23} & {\cellcolor[RGB]{227,193,205} 0.42} & {\cellcolor[RGB]{194,207,220} 0.17} & {\cellcolor[RGB]{184,211,224} 0.10} & {\cellcolor[RGB]{175,214,228} 0.03} & {\cellcolor[RGB]{178,213,227} 0.05} & {\cellcolor[RGB]{232,191,203} 0.46} \\
ICD & {\cellcolor[RGB]{188,209,222} 0.13} & {\cellcolor[RGB]{208,201,213} 0.28} & {\cellcolor[RGB]{202,203,216} 0.23} & {\cellcolor[RGB]{216,197,210} 0.34} & {\cellcolor[RGB]{203,203,216} 0.24} & {\cellcolor[RGB]{188,209,222} 0.13} & {\cellcolor[RGB]{174,215,229} 0.02} & {\cellcolor[RGB]{175,214,228} 0.03} & {\cellcolor[RGB]{241,187,198} 0.53} \\
DRO & {\cellcolor[RGB]{188,209,222} 0.13} & {\cellcolor[RGB]{188,209,222} 0.13} & {\cellcolor[RGB]{207,201,214} 0.27} & {\cellcolor[RGB]{255,182,193} 0.63} & {\cellcolor[RGB]{235,190,201} 0.48} & {\cellcolor[RGB]{190,208,222} 0.14} & {\cellcolor[RGB]{178,213,227} 0.05} & {\cellcolor[RGB]{195,206,219} 0.18} & {\cellcolor[RGB]{232,191,203} 0.46} \\
GoalPriority & {\cellcolor[RGB]{196,206,219} 0.19} & {\cellcolor[RGB]{198,205,218} 0.20} & {\cellcolor[RGB]{210,200,213} 0.29} & {\cellcolor[RGB]{229,192,204} 0.44} & {\cellcolor[RGB]{208,201,213} 0.28} & {\cellcolor[RGB]{196,206,219} 0.19} & {\cellcolor[RGB]{195,206,219} 0.18} & {\cellcolor[RGB]{194,207,220} 0.17} & {\cellcolor[RGB]{232,191,203} 0.46} \\
SmoothLLM & {\cellcolor[RGB]{184,211,224} 0.10} & {\cellcolor[RGB]{194,207,220} 0.17} & {\cellcolor[RGB]{219,196,209} 0.36} & {\cellcolor[RGB]{220,196,208} 0.37} & {\cellcolor[RGB]{224,194,206} 0.40} & {\cellcolor[RGB]{194,207,220} 0.17} & {\cellcolor[RGB]{182,212,225} 0.08} & {\cellcolor[RGB]{183,211,225} 0.09} & {\cellcolor[RGB]{232,191,203} 0.46} \\
SafeDecoding & {\cellcolor[RGB]{199,205,218} 0.21} & {\cellcolor[RGB]{202,203,216} 0.23} & {\cellcolor[RGB]{215,198,210} 0.33} & {\cellcolor[RGB]{235,190,201} 0.48} & {\cellcolor[RGB]{223,195,207} 0.39} & {\cellcolor[RGB]{200,204,217} 0.22} & {\cellcolor[RGB]{200,204,217} 0.22} & {\cellcolor[RGB]{196,206,219} 0.19} & {\cellcolor[RGB]{224,194,206} 0.40} \\
RPO & {\cellcolor[RGB]{190,208,222} 0.14} & {\cellcolor[RGB]{195,206,219} 0.18} & {\cellcolor[RGB]{215,198,210} 0.33} & {\cellcolor[RGB]{227,193,205} 0.42} & {\cellcolor[RGB]{203,203,216} 0.24} & {\cellcolor[RGB]{194,207,220} 0.17} & {\cellcolor[RGB]{196,206,219} 0.19} & {\cellcolor[RGB]{196,206,219} 0.19} & {\cellcolor[RGB]{223,195,207} 0.39} \\
RePE & {\cellcolor[RGB]{191,208,221} 0.15} & {\cellcolor[RGB]{215,198,210} 0.33} & {\cellcolor[RGB]{227,193,205} 0.42} & {\cellcolor[RGB]{239,188,200} 0.51} & {\cellcolor[RGB]{247,185,196} 0.57} & {\cellcolor[RGB]{180,212,226} 0.07} & {\cellcolor[RGB]{195,206,219} 0.18} & {\cellcolor[RGB]{188,209,222} 0.13} & {\cellcolor[RGB]{239,188,200} 0.51} \\
GradSafe & {\cellcolor[RGB]{191,208,221} 0.15} & {\cellcolor[RGB]{195,206,219} 0.18} & {\cellcolor[RGB]{217,197,209} 0.35} & {\cellcolor[RGB]{228,192,204} 0.43} & {\cellcolor[RGB]{228,192,204} 0.43} & {\cellcolor[RGB]{200,204,217} 0.22} & {\cellcolor[RGB]{196,206,219} 0.19} & {\cellcolor[RGB]{190,208,222} 0.14} & {\cellcolor[RGB]{243,186,198} 0.54} \\
AVGAN & {\cellcolor[RGB]{188,209,222} 0.13} & {\cellcolor[RGB]{187,209,223} 0.12} & {\cellcolor[RGB]{199,205,218} 0.21} & {\cellcolor[RGB]{232,191,203} 0.46} & {\cellcolor[RGB]{200,204,217} 0.22} & {\cellcolor[RGB]{187,209,223} 0.12} & {\cellcolor[RGB]{194,207,220} 0.17} & {\cellcolor[RGB]{179,213,227} 0.06} & {\cellcolor[RGB]{220,196,208} 0.37} \\
JBShield & {\cellcolor[RGB]{184,211,224} 0.10} & {\cellcolor[RGB]{221,195,207} 0.38} & {\cellcolor[RGB]{215,198,210} 0.33} & {\cellcolor[RGB]{233,190,202} 0.47} & {\cellcolor[RGB]{239,188,200} 0.51} & {\cellcolor[RGB]{200,204,217} 0.22} & {\cellcolor[RGB]{191,208,221} 0.15} & {\cellcolor[RGB]{191,208,221} 0.15} & {\cellcolor[RGB]{227,193,205} 0.42} \\
RAIN & {\cellcolor[RGB]{196,206,219} 0.19} & {\cellcolor[RGB]{231,191,203} 0.45} & {\cellcolor[RGB]{227,193,205} 0.42} & {\cellcolor[RGB]{202,203,216} 0.23} & {\cellcolor[RGB]{232,191,203} 0.46} & {\cellcolor[RGB]{194,207,220} 0.17} & {\cellcolor[RGB]{198,205,218} 0.20} & {\cellcolor[RGB]{192,207,221} 0.16} & {\cellcolor[RGB]{239,188,200} 0.51} \\
Backdoor Alignment & {\cellcolor[RGB]{199,205,218} 0.21} & {\cellcolor[RGB]{206,202,215} 0.26} & {\cellcolor[RGB]{203,203,216} 0.24} & {\cellcolor[RGB]{196,206,219} 0.19} & {\cellcolor[RGB]{179,213,227} 0.06} & {\cellcolor[RGB]{182,212,225} 0.08} & {\cellcolor[RGB]{200,204,217} 0.22} & {\cellcolor[RGB]{178,213,227} 0.05} & {\cellcolor[RGB]{228,192,204} 0.43} \\
DELMAN & {\cellcolor[RGB]{200,204,217} 0.22} & {\cellcolor[RGB]{237,189,200} 0.50} & {\cellcolor[RGB]{227,193,205} 0.42} & {\cellcolor[RGB]{216,197,210} 0.34} & {\cellcolor[RGB]{227,193,205} 0.42} & {\cellcolor[RGB]{204,202,215} 0.25} & {\cellcolor[RGB]{196,206,219} 0.19} & {\cellcolor[RGB]{184,211,224} 0.10} & {\cellcolor[RGB]{232,191,203} 0.46} \\
Safety-Tuned & {\cellcolor[RGB]{179,213,227} 0.06} & {\cellcolor[RGB]{223,195,207} 0.39} & {\cellcolor[RGB]{203,203,216} 0.24} & {\cellcolor[RGB]{194,207,220} 0.17} & {\cellcolor[RGB]{229,192,204} 0.44} & {\cellcolor[RGB]{192,207,221} 0.16} & {\cellcolor[RGB]{191,208,221} 0.15} & {\cellcolor[RGB]{191,208,221} 0.15} & {\cellcolor[RGB]{228,192,204} 0.43} \\
Layer-AdvPatcher & {\cellcolor[RGB]{175,214,228} 0.03} & {\cellcolor[RGB]{240,188,199} 0.52} & {\cellcolor[RGB]{225,194,206} 0.41} & {\cellcolor[RGB]{233,190,202} 0.47} & {\cellcolor[RGB]{239,188,200} 0.51} & {\cellcolor[RGB]{188,209,222} 0.13} & {\cellcolor[RGB]{232,191,203} 0.46} & {\cellcolor[RGB]{186,210,224} 0.11} & {\cellcolor[RGB]{229,192,204} 0.44} \\
C-advipo & {\cellcolor[RGB]{196,206,219} 0.19} & {\cellcolor[RGB]{211,200,212} 0.30} & {\cellcolor[RGB]{220,196,208} 0.37} & {\cellcolor[RGB]{206,202,215} 0.26} & {\cellcolor[RGB]{210,200,213} 0.29} & {\cellcolor[RGB]{202,203,216} 0.23} & {\cellcolor[RGB]{206,202,215} 0.26} & {\cellcolor[RGB]{191,208,221} 0.15} & {\cellcolor[RGB]{224,194,206} 0.40} \\
\bottomrule
\end{tabular}
    }
    \label{tab:attack_defense_b}
    }
\end{table}
Table~\ref{tab:attack_defense_a} and Table~\ref{tab:attack_defense_b} present a systematic evaluation of various attack-defense combinations on Vicuna-7B. The experimental results demonstrate that defense methods such as RA-LLM and ICD exhibit superior performance against multiple attack variants, effectively reducing ASR and showcasing robust overall defensive capabilities. In contrast, other defense approaches (\textit{e.g.} Paraphrasing and Layer-AdvPatcher) demonstrate limited effectiveness, with ASR remaining at elevated levels. Notably, certain targeted attacks are capable of circumventing specific defense mechanisms, which reveals the limitations in the generalization capabilities of current defense methods and underscores the necessity of developing more robust multi-layered defense architectures.

\subsection{Model-wise Safety Evaluation across Different Risk Categories }

\begin{figure}[tbp]
  \centering
  \begin{minipage}[t]{0.8\textwidth}
    \centering
    \resizebox{\linewidth}{!}{%
    \includegraphics[trim={1.9cm 0.5cm 1.2cm 0.3cm}, clip,width=\textwidth]{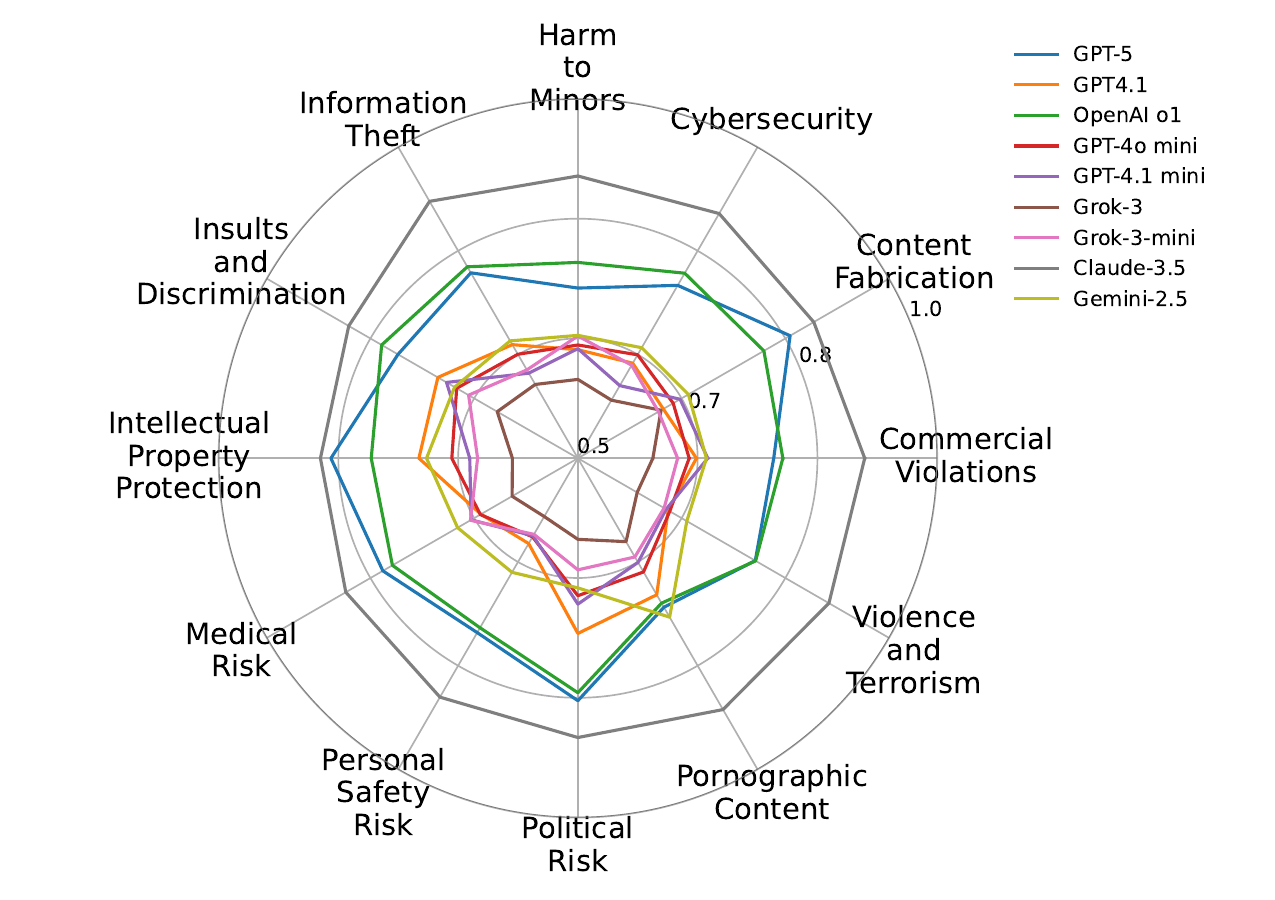}
        }
    \caption{Safety performance of black-box models across different risk categories (using 1-ASR as the metric).}
        \label{fig:radar_black}
  \end{minipage}
  \vfill
  \begin{minipage}[t]{0.8\textwidth}
    \centering
    \resizebox{\linewidth}{!}{%
        \includegraphics[trim={1.9cm 0.5cm 1.2cm 0.3cm}, clip,width=\textwidth]{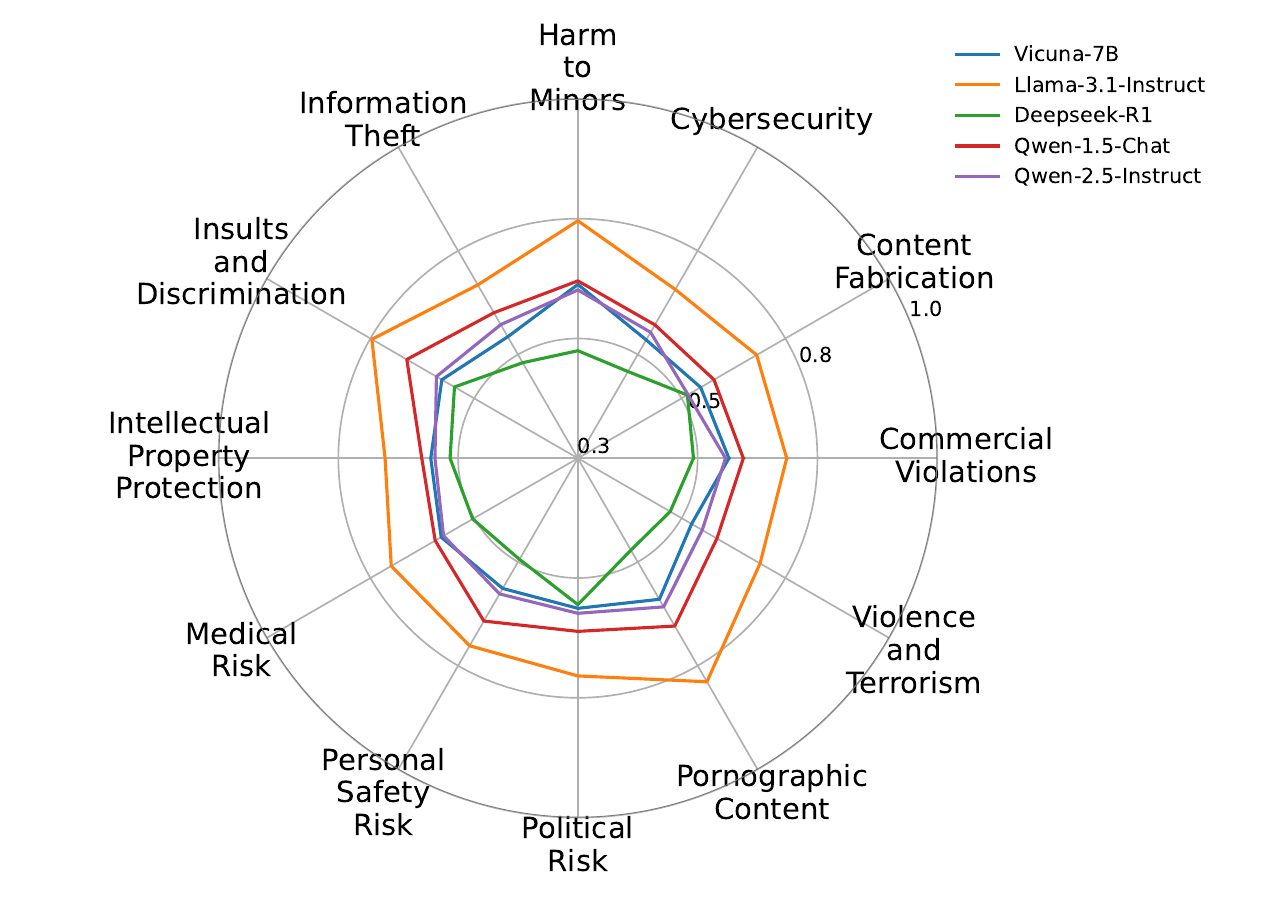}}
    \caption{Safety performance of white-box models across different risk categories (using 1-ASR as the metric).}
        \label{fig:radar_white}
  \end{minipage}
\end{figure}

Based on our framework evaluation, we compute the safety robustness coefficients (defined as $1-\text{ASR}$) for both proprietary and open-source models under evaluation across different risk categories by aggregating results from various attack and evaluation methodologies. The corresponding radar charts are presented in Figure~\ref{fig:radar_black} and Figure~\ref{fig:radar_white}. Among proprietary models, Claude-3.5 demonstrates superior safety defense capabilities across all risk dimensions, followed by GPT-5 and OpenAI o1. Moreover, we observe that most models exhibit relatively consistent defense performance against ``Political Risk'' and ``Pornographic Content'' categories. However, certain models, such as Grok-3 and GPT-4.1 mini demonstrate diminished defense capabilities in the ``Cybersecurity'' dimension. In contrast, open-source models show more uniform performance distributions across all dimensions, with Llama-3.1-Instruct achieving the most robust results. Notably, all open-source models attain relatively high robustness scores in the ``Harm to Minors'', ``Pornographic Content'', and ``Insults and Discrimination'' dimensions. Conversely, adversarial samples targeting synthetic content generation consistently result in elevated ASR values across evaluated models.

\subsection{Discussion}
Our comprehensive evaluation across the TeleAI-Safety framework reveals systematic patterns that extend beyond individual model performance. We discuss three critical implications for the future of LLM safety: the trade-off between reasoning and alignment, the generalization limits of current defenses, and the methodological crisis in automated evaluation.

\paragraph{\textbf{Does Reasoning Compromise Safety}}
A striking observation from our white-box evaluation (Table \ref{tab:black_box}) is the vulnerability of reasoning-specialized models. DeepSeek-R1, explicitly optimized for reasoning via reinforcement learning, exhibits a significantly higher ASR (0.50) compared to general-purpose instruction-tuned models like Llama-3.1 (0.31). We hypothesize that this discrepancy stems from an ``Alignment Tax'' imposed by Chain-of-Thought processes. Advanced attacks, such as DeepInception and Morpheus, often exploit the model's reasoning capabilities by inducing it to generate intermediate steps that ``rationalize'' or ``contextualize'' the harmful request before executing it. This suggests that current safety alignment techniques (\textit{e.g.}, RLHF) may not scale linearly with reasoning depth. As models become more capable of complex logic, they may also become more capable of circumventing their own safety constraints, necessitating a new paradigm of ``Reasoning-Aware Alignment.''

\paragraph{\textbf{The Generalization Crisis in Defensive Mechanisms}} The Generalization Crisis in Defensive Mechanisms Our attack-defense cross-evaluation (Table~\ref{tab:attack_defense_a} and Table~\ref{tab:attack_defense_b}) highlights a ``generalization crisis'' in existing safeguards. We observe that defenses often overfit to specific attack distributions while remaining brittle to others. For instance, perplexity-based filters effectively neutralize optimization-based attacks like GCG (ASR reduced to 0.03), which generate low-probability token sequences. However, these same defenses collapse against semantically coherent, adaptive attacks like Morpheus (ASR 0.33). This indicates that many current defenses rely on surface-level statistical anomalies rather than robust semantic understanding. Consequently, relying on a single defensive layer creates a false sense of security. A robust defense architecture must be multi-layered, combining statistical filtering with semantic analysis to cover the full spectrum of adversarial strategies.

\paragraph{\textbf{The Reliability Gap in Safety Evaluation}} Perhaps the most concerning finding for the research community is the inconsistency of safety evaluators. As evidenced in Table~\ref{tab:eval_attack_part1}, the standard deviation of ASRs across different evaluators for the same attack (\textit{e.g.}, PAIR) reaches as high as 0.34. This variance exposes a fundamental flaw in current ``LLM-as-a-Judge'' methodologies: single-model evaluators suffer from inherent biases, ranging from over-refusal (false positives) to lack of nuance (false negatives). This instability renders cross-paper comparisons nearly impossible and underscores the necessity for consensus-based evaluation frameworks. Our proposed RADAR method, which leverages multi-agent debate to mitigate individual model bias, represents a crucial step toward establishing a more objective and standardized ``Ground Truth'' for LLM safety.
\section{Conclusion and limitations}

In this paper, we present \textbf{TeleAI-Safety}, a comprehensive benchmark and modular framework for evaluating the robustness of LLMs against jailbreak attacks. By integrating 19 attack methods, 29 defensive methods, and 19 evaluation methods within a unified infrastructure, TeleAI-Safety bridges the gap between fragmented safety assessments and systematic benchmarking. Extensive experiments among 17 mainstream models reveal model-specific vulnerabilities, quantify safety–utility trade-offs, and provide an empirical foundation for developing more resilient LLM architectures. In practical scenarios, TeleAI-Safety can be flexibly adjusted with customized attack, defense, and evaluation combinations to meet specific demands.

The key competitiveness of TeleAI-Safety resides in the co-design of its framework and benchmark. This guarantees extensibility, reproducibility, and comprehensiveness across the dimensions of attack, defense, and evaluation. The open-source implementation of TeleAI-Safety allows for continuous updates to adapt to the ever-changing attack and defense strategies. Meanwhile, its standardized evaluation protocol provides a consistent foundation for safety benchmarking within the entire community.

Nevertheless, certain limitations persist. Current evaluations focus primarily on text-based jailbreaks, offering limited insight into multimodal or multilingual vulnerabilities. Moreover, discrepancies among evaluators indicate the need for standardized and consensus-driven scoring schemes. Finally, the computational demands of large-scale benchmarking may limit real-time deployment in constrained environments.

In the future, we aim to extend TeleAI-Safety toward multimodal and cross-lingual safety evaluation, integrate automated red-teaming agents for continual robustness assessment, and explore adaptive defense co-training for dynamic model hardening. We believe that TeleAI-Safety serves a critical role in building transparent, trustworthy, and secure LLM ecosystems for real-world deployments.


\bibliographystyle{elsarticle-num}
\bibliography{refer}

\newpage
\appendix
\begin{center}
    \Large\bf Appendix
\end{center}

\section{Integrated Attacks}  \label{app:app_integrated_attacks}
Contemporary jailbreak attack methodologies can be systematically categorized based on their access level to target models and strategic approaches. From an access perspective, attacks range from white-box methods that leverage complete knowledge of model parameters and architecture to black-box approaches that operate without internal model access. Strategically, the field has evolved from static template-based attacks to sophisticated adaptive optimization methods that employ algorithmic search and semantic transformations.

\textbf{White-box Optimization Attacks.} GCG~\cite{zou2023gcg} pioneered gradient-based optimization of adversarial suffixes by perturbing input tokens using loss gradients as optimization signals. The method maximizes the probability of affirmative prefixes, effectively guiding models to produce unsafe outputs through discrete optimization over token sequences. This approach demonstrated remarkable transferability across different model architectures and established the foundation for subsequent gradient-based methodologies.

\textbf{Gray-box Adaptive Attacks.} AutoDAN~\cite{liu2023autodan} generates jailbreak prompts using hierarchical genetic algorithms, automatically evolving effective attack strategies that bypass model safety mechanisms while maintaining human-readable prompt structures. LAA~\cite{andriushchenko2024laa} designs adaptive templates and appends adversarial suffixes optimized through random search, achieving high success rates by dynamically adjusting attack strategies based on target model responses. AdvPrompter~\cite{paulus2024advprompter} trains specialized attacker language models to autoregressively generate adversarial suffixes for given input queries, enabling automated generation of contextually appropriate jailbreak prompts.

\textbf{Black-box Methods.} GPTFUZZER~\cite{yu2023gptfuzzer} generates new jailbreak templates through iterative mutation of human-written templates, employing five mutation techniques including generation, crossover, expansion, shortening, and rephrasing to systematically explore the prompt space. PAIR~\cite{chao2025pair} utilizes an attacker language model to iteratively refine jailbreak prompts through adaptive red-teaming, enhancing effectiveness through multi-round optimization based on target model feedback.

\textbf{Semantic Transformation Attacks.} These methods preserve malicious intent while modifying linguistic properties to evade safety mechanisms. Past Tense~\cite{andriushchenko2024pasttense} transforms prompts into different grammatical tenses, exploiting temporal framing to circumvent safety constraints by requesting hypothetical or historical scenarios. ArtPrompt~\cite{jiang2024artprompt} disguises harmful requests as artistic or creative endeavors, leveraging ASCII art and visual representations to obfuscate textual instructions. DeepInception~\cite{li2023deepinception} creates diverse fictional scenes and characters to mislead target model safety filters, using nested character structures that collectively work toward harmful goals while bypassing token-level pattern detection.

\textbf{Encoding and Obfuscation Methods.} Cipher~\cite{yuan2023gpt} encodes instructions using various cryptographic schemes such that model safety alignment mechanisms fail to interpret the instructions correctly, enabling jailbreaks through encoding schemes including Caesar ciphers, Base64, and custom encryption formats. MultiLingual~\cite{deng2023multilingual} translates input queries into low-resource languages, exploiting reduced safety mechanism robustness in non-major languages where training data for safety alignment is typically scarce.

\textbf{Structured Search and Tree-based Methods.} TAP~\cite{mehrotra2024tap} maintains a structured tree-form flow to iteratively optimize jailbreak prompts, systematically exploring attack vectors while pruning ineffective branches and expanding promising paths until successful jailbreak prompts are identified.

\textbf{Specialized Strategic Approaches.} Jailbroken~\cite{wei2023jailbroken} targets two key failure modes in LLM safety alignment, including competing objectives and mismatched generalization, crafting prompts that exploit these fundamental weaknesses to bypass safety mechanisms. RENE~\cite{ding2023renellm} combines prompt rewriting and scenario nesting techniques to reframe inputs in ways that circumvent model defenses while maintaining semantic intent. SCAV~\cite{xu2024scav} constructs adversarial prompts strategically by maintaining semantic coherence while introducing carefully designed adversarial variations. ICA~\cite{wei2023ica} employs goal hijacking techniques to redirect model reasoning toward harmful objectives, while Overload~\cite{dong2024overload} exploits token saturation vulnerabilities under high computational load conditions.

Our benchmark further incorporates two cutting-edge attack methodologies developed by our research team: \textbf{Morpheus}, an advanced metacognitive multi-round attack agent.

\section{Integrated Defenses} \label{app:app_integrated_defenses}
Contemporary LLM defense techniques against jailbreak attacks can be systematically classified into two major categories---\textbf{External Defenses} and \textbf{Internal Defenses}---depending on whether they operate at the interface level or are embedded within the model pipeline. Tele-Safety integrates 29 representative defense mechanisms encompassing input/output filtering, semantic regulation, decoding control, knowledge editing, and other integrated safety enhancement techniques.

\textbf{(1) External Defenses.}
External defenses function as protective guardrails surrounding large language models (LLMs), designed to intercept malicious inputs or sanitize potentially unsafe outputs before they reach end users. These approaches are generally \textit{model-agnostic}, enabling seamless integration across diverse architectures and deployment settings.

\paragraph{Input-based Defenses}
Input-based defenses operate by verifying or modifying user prompts before they are processed by the LLM, aiming to intercept harmful queries at an early stage. PPL~\cite{alon2023detecting} evaluates prompt perplexity, where abnormally high or low scores indicate potentially unsafe inputs. Prompt Guard~\cite{meta-llama/Prompt-Guard-86M}, developed by Meta, serves as a specialized classifier to detect malicious prompts, including injection and jailbreak attempts, functioning as a front-end safety filter. Erase and Check~\cite{kumar2023certifying} and RA-LLM~\cite{cao2024defending} both adopt token-level verification to enhance input safety, but they differ in strategy. Erase and Check systematically enumerates sub-sequences of a prompt and verifies each for potential harm, ensuring that the original input cannot trigger unsafe outputs. RA-LLM, in contrast, applies random deletions to the input and evaluates the model’s responses to these perturbations, using the resulting behavioral consistency to determine whether the prompt is safe. Paraphrasing~\cite{jain2023baseline} reformulates user inputs into simpler or more neutral expressions, mitigating the risk of triggering unsafe behaviors. SmoothLLM~\cite{robey2023smoothllm} generates diverse perturbations of prompts at the character level and aggregates the model’s responses to detect and weaken the influence of adversarial inputs, thereby enhancing robustness against jailbreak attacks. 

Beyond linguistic transformation, more advanced frameworks leverage semantic understanding and information theory. IBProtector~\cite{liu2024protecting} employs the information bottleneck principle to selectively compress and perturb prompts, retaining only information essential for generating intended responses while filtering out harmful or irrelevant details. EDDF~\cite{xiang2025beyond} extracts and stores the ``essence'' of known jailbreak attacks during an offline stage and subsequently detects whether new prompts contain these adversarial characteristics during inference, enabling real-time malicious prompt identification. BackTranslation~\cite{wang2024defending} infers the true intent of prompts by reverse-generating potential source prompts from preliminary model outputs, exposing concealed malicious goals and allowing the system to reject unsafe inputs before they are processed by the main model.

\paragraph{Output-based Defenses}
Output-based defenses operate after the model generates responses, aiming to detect, filter, or revise unsafe content before it reaches end users. These methods serve as a post-hoc safety layer that ensures harmful outputs are intercepted or corrected. A straightforward approach is Self-Defense~\cite{phute2023llm}, which reuses the model itself as a safety verifier. After an initial response is produced, the model re-evaluates its own output to identify potentially harmful or policy-violating content, effectively acting as a self-checking mechanism. Building upon this paradigm, Aligner~\cite{ji2024aligner} introduces a lightweight, plug-and-play module that reviews and adjusts the main LLM’s outputs. By integrating an auxiliary alignment model, Aligner automatically corrects unsafe or misaligned responses before they are delivered to the user. To further enhance interpretability and precision, GuardReasoner~\cite{liu2025guardreasoner} combines step-by-step reasoning with preference-based optimization, constructing reasoning chains that explain why a response might be unsafe and refining outputs through a structured, multi-stage evaluation process.

\textbf{(2) Internal Defenses.}
Internal defenses aim to strengthen the model itself against jailbreak attacks by modifying or constraining its internal behavior during inference or training. Unlike external defenses that act at the interface level, internal strategies directly enhance the model’s robustness and alignment through controlled optimization or architectural interventions.

\paragraph{Inference-time Defenses}
Inference-time defenses aim to regulate the model’s behavior during response generation by introducing safety constraints or contextual guidance, without modifying its underlying parameters. Prompt-level control includes methods that modify the system prompt or embed safety demonstrations to guide safe reasoning. SelfReminder~\cite{xie2023defending} inserts explicit safety reminders into the model’s internal context, maintaining continuous awareness of alignment principles. GoalPriority~\cite{zhang2024defending} reformulates system instruction to explicitly prioritize safety objectives over helpfulness, promoting conservative outputs when conflicts arise. In-context learning–based approaches embed safety demonstrations into the prompt to guide model behavior at runtime. ICD~\cite{wei2023jailbreak} leverages few-shot examples that illustrate aligned behaviors, enabling the model to internalize safe response patterns through implicit contextual learning. RPO~\cite{zhou2024robust} refines prompt suffixes through robust optimization, enhancing resistance to adversarial inputs.

Internal representation–based approaches intervene within hidden layers to detect or suppress unsafe behavior. RePE~\cite{zou2023representation} identifies “concept directions” in the hidden layers that correspond to safety-relevant semantics. By selectively amplifying or suppressing these directions, RePE can directly reduce the model’s tendency to produce harmful outputs while keeping normal responses unaffected. DRO~\cite{zheng2024prompt} encodes safety instructions as latent vectors and applies robust optimization to guide hidden representations, ensuring harmful inputs trigger refusal while benign inputs remain unaffected. JBShield~\cite{ji2024aligner} monitors activations corresponding to harmful or jailbreak concepts, attenuating unsafe signals while reinforcing safe responses. AVGAN~\cite{li2025cavgan} uses a GAN framework in representation space to learn the boundary between safe and unsafe activations.

Gradient-oriented approaches analyze the model’s gradient dynamics to detect potential jailbreak attempts. GradSafe~\cite{xie2024gradsafe} compares gradient changes of safety-critical parameters to known malicious patterns, enabling zero-shot detection of harmful prompts. Gradient Cuff~\cite{hu2024gradient} defines a “refusal loss” to measure the model’s tendency to follow unsafe instructions and monitors both its value and gradient norm to detect jailbreak-like inputs while maintaining normal performance.

Output-focused methods further enhance safety by intervening in the decoding process. SafeDecoding~\cite{xu2024safedecoding} identifies safety tokens during decoding and adjusts token probabilities to suppress harmful sequences, leveraging safety token ranking to mitigate attacks. RAIN~\cite{li2023rain} enables the model to self-evaluate ongoing outputs and revert to an earlier safe state when necessary, implementing self-correction without requiring extra alignment data or fine-tuning.

\paragraph{Training-time Defenses}
Training-time defenses aim to enhance the model’s inherent safety by adjusting parameters during training or fine-tuning. Unlike inference-time defenses, training-time methods adjust the model’s parameters during training to internalize safety behaviors.

Fine-tuning and RLHF-based alignment guide the model toward safe responses using curated data or human feedback. Safety-Tuned LLaMAs~\cite{bianchi2024safety} incorporate safety-oriented examples during instruction tuning to bias the model toward harmless outputs.

Knowledge and model editing methods directly modify model parameters or layers to remove vulnerabilities. DELMAN~\cite{wang2025delman} edits minimal sets of parameters associated with harmful behaviors, using KL regularization to preserve normal outputs while adapting to new attacks. Layer-AdvPatcher~\cite{ouyang2025layer} targets layers prone to generating affirmative tokens and applies self-enhanced training to reduce harmful responses without affecting normal queries.

\section{Integrated Evaluations}  
\label{app:app_integrated_evaluations}
Contemporary LLM safety evaluation methods can be vertically classified into multiple categories based on the semantic understanding capabilities required by the evaluators. Benefiting from significant advances in LLM language comprehension abilities, safety evaluation approaches demonstrate an overall trend of transitioning from reliance on traditional rule-based matching toward semantic content-based risk prevention mechanisms. This paper integrates 18 established classical evaluation methods along with 1 novel evaluation technique, encompassing four core mechanism categories: rule-based, fine-tuned llm-based, chat LLM-based, and multi-agent system-based approaches, to comprehensively address various safety challenges.

\textbf{Rule-based evaluators.}
Rule-based evaluation methods rely on predefined pattern matching and keyword lists for rapid screening to efficiently identify known malicious query patterns. For instance, AISafetyLab~\cite{zhang2025aisafetylab} integrates safety evaluators primarily based on prefix matching and pattern matching, enabling quick identification of potentially harmful content through static rules. In jailbreaking attack research, Harmbench~\cite{mazeika2024harmbench} and GptFuzzer~\cite{yu2023gptfuzzer}, in addition to introducing jailbreaking attack methods, also contribute corresponding language classification models for rule-based harmful content identification and classification.

\textbf{Fine-tuned llm-based evaluators.}
Fine-tuned LLM-based evaluation methods employ specially trained language models for risk classification and control, enabling models to provide judgment conclusions and analysis for jailbreaking attack cases through fine-tuning techniques. Examples include \texttt{ShieldLM}~\cite{zhang2024shieldlm}, \texttt{LlamaGuard3}, and \texttt{ShieldGemma}~\cite{zeng2024shieldgemmagenerativeaicontent}, which serve as risk controllers that make risk-level decisions regarding attack-defense outcomes. Meanwhile, the classification models in \texttt{Harmbench}~\cite{mazeika2024harmbench} and \texttt{GptFuzzer}~\cite{yu2023gptfuzzer} also belong to this category, achieving high-precision harmful content identification through fine-tuning.

\textbf{Chat LLM-based evaluators.}
Chat LLM-based evaluation methods leverage the powerful language understanding capabilities of general-purpose conversational models to accomplish risk assessment tasks through specifically designed prompt templates or paradigms. For example, risk evaluators such as \texttt{QwenScorer}, \texttt{GPT5Scorer}, and \texttt{ClaudeScorer} employ such methods to perform risk judgment on LLM outputs. Additionally, methods proposed in~\cite{shu2025attackeval, qi2023fine,mehrotra2024tree,chao2025jailbreaking,zhang2025aisafetylab, zhang2024shieldlm} achieve efficient evaluation through single-model interaction by optimizing prompt templates and interaction paradigms, thereby enhancing the accuracy of semantic understanding.

\textbf{Multi-agent Collaboration-based evaluators.}
Multi-agent Collaboration-based evaluation methods integrate multiple specialized agents working collaboratively to enhance detection coverage and reliability through multi-round deliberation or consensus mechanisms. For instance, the \texttt{RADAR}~\cite{chen2025radar} framework simulates multi-round debates among multiple evaluators to reach consensus judgments regarding query risks and model response safety. This approach effectively addresses increasingly complex safety challenges and reduces bias from individual evaluators through multi-agent collaboration.

\section{Usage}
\subsection{Attack Implementation}
TeleAI-Safety provides a unified and user-friendly interface for executing jailbreak attacks through a command-line-driven approach. Each attack method can be launched independently using a standardized execution pattern that ensures consistency and reproducibility across all integrated techniques.
To execute an attack, users invoke the corresponding attack script with a YAML configuration file that specifies all necessary parameters:
\begin{tcblisting}{%
    listing only,
    listing options={language=bash, basicstyle=\ttfamily\small},
    % colback=gray!5,
    colframe=gray!30,
    boxrule=0.5pt,
    arc=3pt,
    left=6pt,
    right=6pt,
    top=3pt,
    bottom=3pt,
    fontupper=\small
}
python <attack_name>.py --config_path=./configs/<attack_name>.yaml
\end{tcblisting}
For example, to run the PAIR attack method, the command would be:
\begin{tcblisting}{%
    listing only,
    listing options={language=bash, basicstyle=\ttfamily\small},
    % colback=gray!5,
    colframe=gray!30,
    boxrule=0.5pt,
    arc=3pt,
    left=6pt,
    right=6pt,
    top=3pt,
    bottom=3pt,
    fontupper=\small
}
python pair.py --config_path=./configs/pair.yaml
\end{tcblisting}

The configuration file follows a flat key-value structure that encompasses four primary components: dataset specification, target model configuration, API/URL credentials, and attack-specific hyperparameters. A typical configuration file structure is illustrated below:
\begin{tcblisting}{%
    listing only,
    listing options={language=bash, basicstyle=\ttfamily\small},
    % colback=gray!5,
    colframe=gray!30,
    boxrule=0.5pt,
    arc=3pt,
    left=6pt,
    right=6pt,
    top=3pt,
    bottom=3pt,
    fontupper=\small
}
# Dataset Configuration
dataset_path: ./data/attack_dataset.json

# Target Model Configuration
target_model_type: local  # Options: openai, local
target_model_name: vicuna-7b-v1.5
target_model_path: lmsys/vicuna-7b-v1.5   # For local models only

# API Configuration
api_key: ${OPENAI_API_KEY}
base_url: ${OPENAI_BASE_URL}

# Results Configuration
results_path: ./results/<results_name>.jsonl

# Attack-Specific Parameters
temperature: 0.7
max_tokens: 512
\end{tcblisting}





The \texttt{dataset\_path} field specifies the path to the attack corpus containing the evaluation samples. The \texttt{target\_model\_type} field indicates whether the victim model is accessed via API or loaded locally, with \texttt{target\_model\_name} and \texttt{target\_model\_path} specifying the model identifier and local checkpoint path, respectively. API credentials are provided through \texttt{api\_key} and \texttt{base\_url} fields, supporting both commercial services and self-hosted endpoints. The \texttt{results\_path} field defines the output directory for storing attack logs and evaluation metrics. Additional fields contain method-specific configurations that vary across different attack techniques, such as the number of optimization iterations for gradient-based attacks or the tree search depth for methods like TAP.
For attacks requiring auxiliary models, such as PAIR which uses an attacker LLM to generate adversarial prompts, additional configuration fields can be specified:
\begin{tcblisting}{%
    listing only,
    listing options={language=bash, basicstyle=\ttfamily\small},
    % colback=gray!5,
    colframe=gray!30,
    boxrule=0.5pt,
    arc=3pt,
    left=6pt,
    right=6pt,
    top=3pt,
    bottom=3pt,
    fontupper=\small
}
# Auxiliary Model Configuration (for attacks like PAIR)
attack_model_type: local
attack_model_name: vicuna-13b-v1.5
attack_model_path: ./models/vicuna-13b-v1.5
device: cuda:0
batch_size: 8
\end{tcblisting}

All attack implementations automatically log results to a structured output directory, capturing both successful and unsuccessful jailbreak attempts along with relevant metadata such as ASR, query efficiency, and response time. This standardized logging format facilitates downstream analysis and cross-method comparison, enabling researchers to systematically evaluate attack performance across different target models and defensive configurations.

\subsection{Defense Implementation}
Building upon the unified structure for attack methods, TeleAI-Safety also offers a command-line interface for executing defense mechanisms against various attacks. The defense framework is designed to be flexible and reproducible, allowing users to apply a range of defense techniques to protect target models. Just like the attack methods, each defense strategy can be invoked independently, ensuring consistency across different types of defenses. Furthermore, multiple defense strategies can be applied simultaneously, providing enhanced protection against diverse attack vectors. 

To execute defense, users invoke the corresponding defense script with two YAML configuration files: 
\texttt{defender\_config} and \texttt{filter\_config}. The \texttt{defender\_config} file specifies parameters that are directly related to the defense strategy, while the \texttt{filter\_config} governs general execution settings. The command structure is as follows:
\begin{tcblisting}{%
    listing only,
    listing options={language=bash, basicstyle=\ttfamily\small},
    % colback=gray!5,
    colframe=gray!30,
    boxrule=0.5pt,
    arc=3pt,
    left=6pt,
    right=6pt,
    top=3pt,
    bottom=3pt,
    fontupper=\small
}
python run_defense.py --defender_config=./configs/defender.yaml --filter_config=./configs/filter.yaml
\end{tcblisting}

The defender configuration file specifies parameters directly related to the defense strategy. It defines the type of defense, model settings, defense-specific hyperparameters, and other parameters that control the behavior of the defense mechanism. This configuration serves as the core setup for executing the defense.

The filter configuration file is used for non-defender parameters that influence the general execution of the defense process. It specifies batch sizes, logging settings, target model names, and attack data paths, among others. This file allows users to control how the defense interacts with different datasets, allocate resources like GPUs, and adjust other parameters that affect the batch processing of defense operations.
\begin{tcblisting}{%
    listing only,
    listing options={language=bash, basicstyle=\ttfamily\small},
    % colback=gray!5,
    colframe=gray!30,
    boxrule=0.5pt,
    arc=3pt,
    left=6pt,
    right=6pt,
    top=3pt,
    bottom=3pt,
    fontupper=\small
}
# Attack Configuration
attack_data_path: ./data/attack_samples.jsonl
attack_types:
  - "ica"
  - "cipher"
  - "jailbroken"
target_model: "vicuna-7b-v1.5"
# Batch Size
batch_size: 8
# Log Level
log_level: INFO
# Results Storage Directory
save_results_dir: ./defense_results
\end{tcblisting}

By separating the configurations into two distinct files, TeleAI-Safety offers enhanced flexibility, allowing users to adjust the defense parameters and execution settings independently. This separation ensures streamlined and efficient operation across different defense strategies, providing users with both customization and consistency in their defense processes.

\subsection{Evaluation Implementation}
TeleAI-Safety integrates multiple out-of-the-box evaluators for assessing the outcomes generated by attacks. Similar to the usage of attack modules, each evaluator can be independently launched using standardized configuration files defined in YAML format.

To execute the evaluation, users need to invoke the evaluation script with a YAML configuration file and the results to be evaluated. The configuration file contains the necessary parameters for the evaluator, while the results to be evaluated include the responses generated by the model after attacks. Similarly, the command for executing the evaluation is as follows:

\begin{tcblisting}{%
    listing only,
    listing options={language=bash, basicstyle=\ttfamily\small},
    % colback=gray!5,
    colframe=gray!30,
    boxrule=0.5pt,
    arc=3pt,
    left=6pt,
    right=6pt,
    top=3pt,
    bottom=3pt,
    fontupper=\small
}
python eval_summary_report.py --scorer=<evaluator_name> \
    --config_path=<path_to_config_file> \
    --json_path=<path_to_attack_result> \
    --output_path=<path_to_output_file>
\end{tcblisting}

The configuration file adopts a flat key-value structure, primarily containing weight paths, device settings, and role configurations for multi-agent collaboration related to RADAR. The output file records statistical metric reports of the evaluation, as shown below:
\begin{tcblisting}{%
    listing only,
    listing options={language=bash, basicstyle=\ttfamily\small},
    % colback=gray!5,
    colframe=gray!30,
    boxrule=0.5pt,
    arc=3pt,
    left=6pt,
    right=6pt,
    top=3pt,
    bottom=3pt,
    fontupper=\small
}
Evaluation Summary Report
Scorer: PatternScorer
Input file: xxx
Total samples: 342
Attack success samples: 63
Attack Success Rate (ASR): 0.1842
\end{tcblisting}

\end{document}